\documentclass[useAMS,usenatbib]{mn2e}
\usepackage{aas_macros}
\usepackage[a4paper,centering, totalwidth=520pt, totalheight=700pt]{geometry}
\usepackage{graphicx}
\usepackage{amssymb}
\usepackage{amsmath}
\usepackage{enumerate}
\usepackage{microtype}

\usepackage{color}

\newcommand{\Xv}{ {\bf X}}

\newcommand{\p}{\ensuremath{\partial}}
\newcommand{\del}{\ensuremath{\delta}}
\newcommand{\Del}{\ensuremath{\Delta}}
\newcommand{\lam}{\ensuremath{\lambda}}
\newcommand{\gam}{\ensuremath{\gamma}}

\newcommand{\sig}{\ensuremath{\sigma}}

\newcommand{\delc}{\ensuremath{\delta_{\rm c}}}

\newcommand{\Ms}{\ensuremath{\,h^{-1}{\rm M}_{\odot}}}

\newcommand{\avg}[1]{\ensuremath{\left\langle \,#1\, \right\rangle}}
\newcommand{\nab}{\ensuremath{\boldsymbol{\nabla}}}

\newcommand{\der}{\ensuremath{{\rm d}}}
\newcommand{\dir}{\ensuremath{\delta_{\rm D}}}

\newcommand{\erfc}[1]{\ensuremath{{\rm erfc}\left(#1\right)}}
\newcommand{\erf}[1]{\ensuremath{{\rm erf}\left(#1\right)}}

\newcommand{\eqn}[1]{equation~\eqref{#1}}
\newcommand{\eqns}[1]{equations~\eqref{#1}}
\newcommand{\fig}[1]{Figure~\ref{#1}}

\newcommand{\ph}[1]{\phantom{#1}}

\newcommand{\be}{\begin{equation}}
\newcommand{\ee}{\end{equation}}
\newcommand{\Cal}[1]{\ensuremath{\mathcal{#1}}}

\def \peaks {ESPeaks}
\def \peak {ESPeak}

\title[Locations of halo formation]{The locations of halo formation and the peaks formalism}
\author[O. Hahn \& A. Paranjape]{Oliver Hahn\thanks{Email:hahn@phys.ethz.ch}$^1$ and 
Aseem Paranjape\thanks{Email:aseemp@phys.ethz.ch}$^1$ \\
$^{1}$Institute for Astronomy, Department of Physics, ETH Zurich, CH-8093 Z\"urich, Switzerland
  }
  
\begin{document}
\date{MNRAS in press}
\pagerange{\pageref{firstpage}--\pageref{lastpage}} \pubyear{2013}
\maketitle
\label{firstpage}

\begin{abstract}
We investigate the problem of predicting the halo mass function from the properties of the Lagrangian density field. We focus on a perturbation spectrum with a small-scale cut-off (as in warm dark matter cosmologies). This cut-off results in a strong suppression of low mass objects, providing additional leverage to rigorously test which perturbations collapse and to what mass. We find that all haloes are consistent with forming near peaks of the initial density field, with a strong correlation between proto-halo density and ellipticity. We demonstrate that, while standard excursion set theory with correlated steps completely fails to reproduce the mass function, the inclusion of the peaks constraint leads to the correct number of haloes but significantly underpredicts the masses of low-mass objects (with the predicted halo mass function at low masses behaving like $\der n/\der\ln m \sim m^{2/3}$). This prediction is very robust and cannot be easily altered within the framework of a single collapse barrier. The nature of collapse in the presence of a small-scale cut-off thus reveals that excursion set calculations require a more detailed understanding of the collapse-time of a general ellipsoidal perturbation to predict the ultimate collapsed mass of a peak -- a problem that has been hidden in the large abundance of small-scale structure in CDM. We demonstrate how this problem can be resolved within the excursion set framework.
\end{abstract}



\section{Introduction}
\label{sec:intro}

Where and when do dark matter haloes form?
The problem of identifying the locations where gravitational collapse leads to bound haloes of dark matter, and predicting the cosmic time at which this will occur, is among the oldest problems of cosmic structure formation theory. 
The idea that small perturbations in the primordial matter density formed the seeds of the large-scale structure we observe at present is among the cornerstones of our current picture of the evolution of the Universe. An understanding of the relevant processes and a robust theoretical model enables us to map properties such as the abundance and clustering of dark matter haloes -- which are directly tied to the corresponding observed properties of galaxies -- to well-understood statistical properties of the \emph{initial} dark matter density. 

Although this problem can now be tackled directly using numerical simulations of large, cosmological volumes, it is still important to explore analytical approximations and identify the key physical features that decide the sites of halo formation. The main motivation behind this exercise is to gain a better understanding of the physical processes that affect structure formation in the Universe. From a practical viewpoint, however, this can also lead to useful, fast approximations to the halo mass function, clustering, and predictions of collapse time for a given patch in the initial conditions. The latter especially could be useful from the point of view of ``semi-analytic'' mock catalog algorithms such as PTHALOS \citep{ss02}, {\sc Pinocchio} \citep{pinocchio,pinocchio-reloaded}, COLA \citep*{tze13}, ALPT \citep{kh12}, etc., which are becoming increasingly popular in the construction of covariance matrices in current and upcoming suveys such as BOSS \citep{Manera13}, WiggleZ \citep{Marin13}, Euclid \citep{Laureijs11}, and others. 

Our focus in this paper is on the mass function of dark matter haloes, which is the most basic diagnostic of the fully non-linear density field. Analytical descriptions of the halo mass function have traditionally used two parallel approaches: the excursion set approach \citep{ps74,e83,ph90,bcek91,lc93,s98,smt01,mr10,pls12,ms12,arsc12,ms13} and the peaks formalism \citep{bbks86,Bond1989,aj90,m+98,h01}, both of which aim to characterize the locations of collapse in the initial conditions using some criteria. The former relies on counting sufficiently overdense regions in the initial conditions, which it maps to collapsed haloes in the final, gravitationally evolved density field, while the latter associates haloes specifically to peaks in the initial matter density. In other words, while both approaches rely on the statistical properties of the initial conditions to predict final halo abundances, the excursion set approach does this by treating all locations in the initial conditions on the same footing, while the peaks formalism treats density peaks as being special. 

The key aspect of the excursion set approach \citep{bcek91}, which is missing in the traditional peaks approach \citep{bbks86}, is that it explicitly accounts for the so-called ``cloud-in-cloud'' problem which avoids overcounting overdense regions embedded in larger overdense regions as individual objects. The ``peak-patch'' approach of \cite{bm96} is a numerical prescription for unifying the two approaches to solve the cloud-in-cloud problem for peaks, or, equivalently, to study excursion sets for a special subset of initial locations, namely peaks. Recent work \citep{ms12,ps12} has shown that this can also be achieved analytically by making some simple but accurate approximations \citep[see also][]{Bond1989}. There are several motivations for doing so \citep*{smt01,ps12}, not least the fact that $N$-body simulations of cold dark matter (CDM) show that a large fraction of haloes do, in fact, originate from initial density peaks \citep{lp11}. Further, \citet*{psd13} showed that this unified analytical formalism of excursion set peaks (ESP) gives a self-consistent description of the CDM halo mass function as well as clustering which is accurate at the $\sim10\%$ level.

It is worth asking whether this formalism has correctly captured all the relevant aspects of structure formation that affect the mass function. One way of addressing this issue is to apply the same formalism in an ``extreme'' situation which it was not explicitly built to describe. Structure formation from an initial matter power spectrum with highly suppressed small-scale power, as found in warm dark matter (WDM) cosmologies, offers the perfect playground. The reason is that, apart from having a truncated initial power spectrum, \emph{simulations of WDM in fact solve exactly the same problem as those of CDM}: 
the evolution of a cold, collisionless, self-gravitating fluid. 

Analytically, one then expects that the same ESP expressions, which correctly describe the CDM mass function and clustering, should work for the WDM case as well, with the simple replacement of the CDM initial power spectrum with that of WDM. In this regard, as we describe in detail below, the ``out-of-the-box'' ESP calculation does considerably better than traditional TopHat-filtered excursion sets: it correctly predicts a turnover in $\der n/\der\ln m$ at the correct scale whereas the latter predicts a monotonic rise at low masses. 
We will see, however, that ESP predicts a power law decrease at low masses $\der n/\der\ln m \sim m^{2/3}$ which is incompatible with the results of simulations. This analytical prediction is very robust and hints at a missing physical ingredient in the excursion set logic\footnote{We should note that previous authors \citep[e.g.,][]{b+13,ssr13} have motivated a standard excursion set analysis of the WDM mass function (without the peaks constraint) by appealing to a smoothing filter that is sharp in Fourier space. While the resulting mass function fits are straightforward to implement, the physical relevance of the sharp-$k$ filter is less clear. Although there might be a deeper reason behind its success (e.g., it could be that the real-space nonlocality inherent in the sharp-$k$ filter somehow captures the properties of the initial density environment near small mass WDM peaks better than, say, the TopHat filter), we believe it is important to first assess how well the physically motivated picture of peaks itself fares. We will therefore not pursue sharp-$k$ filtering in this paper.}.

Our goal in this paper is to characterise the collapsed objects identified in a WDM simulation in terms of the properties of the initial density field. This will allow us to understand the reasons behind the mismatch of the measured mass function and the ESP prediction. The paper is structured as follows: 

In Section~\ref{sec:massfunc}, we describe the numerical simulation and halo finding algorithm, which are the same as presented by \citet*{Angulo2013}. We then compare the resulting halo mass function with theoretical expectations based on the ESP formalism, and discuss possible reasons for the differences we see between the theory and numerics. To better understand where haloes form, in Sections~\ref{sec:haloprops} and~\ref{sec:empiricalwalks} we turn to an in-depth analysis of the initial conditions of the simulation. In Section~\ref{sec:haloprops} we analyse the initial density field at the ``Lagrangian patches'' of the haloes (i.e., the initial locations of groups of particles that will eventually be identified as haloes) and demonstrate that all haloes in the simulation are consistent with forming near peaks of the initial density. We also explore correlations between the initial overdensity and shape of the Lagrangian patches, and use these results to motivate the construction of an empirical catalogue of ``\peaks'', which we describe in Section~\ref{sec:empiricalwalks}. These \peaks\ are a numerical realisation of what the ESP calculation aims to accomplish, and we compare their Lagrangian properties with those of the haloes. 

Our main conclusion from this exercise is that, while the ESP calculation on average correctly identifies the locations of halo formation, it systematically underpredicts the mass of the resulting object, and that this effect is especially enhanced at halo masses that are small compared to the characteristic mass scale where the WDM mass function turns around. In Section~\ref{sec:analytical} we argue that this mass mismatch is related to a systematic overprediction of the \emph{time} of collapse of a given perturbation, and propose a modification to the ESP calculation to re-assign masses by correcting for this effect. We show that the resulting mass function not only agrees very well with the WDM result, but also describes the \emph{CDM} mass function accurately with the simple replacement of the WDM power spectrum with that of CDM.

We close with a summary and discussion in Section~\ref{sec:discussion}. The Appendices collect technical details and arguments used to motivate some of the results in the main text.


\section{The Halo Mass Function: Confronting simulations and theory}
\label{sec:massfunc}

Matter power spectra with an initial small-scale truncation
arise naturally in warm and hot dark matter cosmologies where density fluctuations on small scales are suppressed due to the late transition to the non-relativistic regime of the respective dark matter particle. Such a power spectrum leads to a corresponding turn-over in the late time halo mass function. The numerical determination of such mass functions has, however, proven extremely challenging due to the presence of low mass objects that arise -- completely unphysically -- from the fragmentation of filaments ~\citep[see e.g.][]{Avila-Reese:2001,Bode2001,Wang2007,Melott2007,Hahn2013}. In the presence of artificial fragmentation, mass functions can only be measured indirectly after filtering or correcting for the spurious haloes \citep[see e.g.][]{Lovell2012,Schneider2012,Lovell2013}. Only more recently has the behaviour of the halo mass function around and below the turn-over scale been explicitly demonstrated by \citet[][AHA13, in what follows]{Angulo2013}. 

While such WDM cosmologies are of course of genuine physical interest in their own right, we are mainly concerned with a different aspect here: the suppression of \emph{low} mass haloes provides powerful additional leverage to test models of structure formation in such cosmologies. The exponential fall in the halo mass function at large masses -- whose sensitivity to cosmological parameters has been exploited for decades -- is replicated here at the small-mass end. Any theoretical model must now describe both of these strong features in the mass function.

We begin by briefly discussing the numerical simulations of AHA13 and the WDM halo mass function they measure. We will see that these numerical results do not meet theoretical expectations based on the ESP formalism. We discuss possible reasons for this, which will motivate our subsequent analysis.


\subsection{Numerical Simulation}
The numerical simulation discussed by AHA13 employs the novel T4PM method \citep{Hahn2013}, which completely suppresses artificial fragmentation and allows the determination of the halo mass function at and below the turn-over scale in the absence of numerical artefacts.

Specifically, this simulation resolves a $80\,h^{-1}{\rm Mpc}$ cosmological volume with $1024^3$ particles with cosmological parameters $\Omega_{\rm m}=0.276$, $\Omega_\Lambda=0.724$, $\Omega_{\rm b}=0.045$, $h=0.703$, $\sigma_8=0.811$ and $n_{\rm s}=0.96$, consistent with the WMAP7 data release \citep{Komatsu2010}. The normalisation of the power spectrum using $\sigma_8$ was set using a CDM spectrum, so that the amplitude of fluctuations on large-scales is independent of the truncation scale of the power spectrum.

The truncation of power at small scales is done by assuming a toy model cosmology with a $0.25\,{\rm keV}$ thermally produced WDM particle. Such a particle is, of course, completely ruled out by observations as the dominant component of dark matter \citep[see e.g.][who derive a current lower bound of 3.3~keV]{Viel2013}. However, it allows resolving the entire power spectrum up to the truncation scale with sufficient particles and, as we have already argued above, is studied in this paper for the main purpose of testing analytical predictions. In particular, AHA13 used the fitting formula of \cite{Bode2001} to modify the CDM transfer function
\begin{equation}
 T_{\rm WDM}(k) = T_{\rm CDM}(k) \left[1 + (\alpha\,k)^2\right]^{-5.0},
\label{Tk-WDM}
\end{equation}
with
\begin{equation}
\alpha  \equiv 0.05 \left(\frac{\Omega_m}{0.4}\right)^{0.15} \left(\frac{h}{0.65}\right)^{1.3}
\left( \frac{m_{\rm dm}}{1\,{\rm keV}}  \right)^{-1.15}\,h^{-1}{\rm Mpc}, 
\label{alpha}
\end{equation}
where $m_{\rm dm}(=0.25{\rm keV})$ is the DM particle mass. 
This results in $\alpha = 0.26\,h^{-1}{\rm Mpc}$, equivalent to a free-streaming mass-scale 
\begin{equation}
M_{\rm fs} =\frac{4\pi}{3}\bar{\rho}\left(\alpha/2\right)^3\simeq7\times10^{8}\Ms\,,
\label{Mfs}
\end{equation}
and a ``half-mode'' mass-scale \citep[c.f., e.g.,][]{Schneider2012}
\begin{equation}
M_{\rm hm}\simeq 4.3\times10^{3}\,M_{\rm fs}\simeq3.0\times10^{12}\Ms.
\label{Mhm}
\end{equation}
Note that, as discussed in more detail in AHA13, these simulations do not include the (small) thermal velocity dispersion that a real WDM fluid would possess, so that the collisionless dark matter fluid is in fact treated in the perfectly cold limit, after perturbations have been suppressed below the maximum free-streaming scale in linear perturbation theory. A thermal velocity however is expected to have little effect on the abundance of collapsed structures at late times, which is the main topic of our interest here.

Adopting the fit of \cite{EisensteinHu1999} as the fiducial CDM transfer function $T_{\rm CDM}$, initial conditions were generated using the {\sc Music} code \citep{Hahn2011} at an initial redshift of $z=63$ using the Zel'dovich approximation. We note that a simulation initialized at such a rather low redshift using first order Lagrangian perturbation theory is to some degree affected by transients from the initial conditions \citep[e.g. ][]{Crocce2006}. The high-mass end of the halo mass function is thus expected to deviate from the true one. The small volume of the simulation adds further to a systematic deviation. Furthermore, it is possible that the detailed behaviour of the mass function around the half-mode mass is also affected by transients. This possibility needs to be considered for precision determinations of the halo mass function but we do not expect it to alter the qualitative behaviour with which we are mostly concerned here (the results of AHA13 are roughly consistent with, e.g., the predictions of \citealt{ssr13} who use 2LPT). We note that the half-mode mass is resolved with almost $100'000$ particles in the simulation of AHA13.


\subsection{Halo Identification}
\label{sec:halofinding}
AHA13 found that the suppression of artificial fragmentation leads to a failure of the Friends-of-Friends (FoF) halo finder. The dense cores of filaments (in the absence of artificial fragmentation) lead to a percolation of large regions of several haloes when the standard linking parameter $b=0.2$ is used. Instead, they first adopted a linking parameter of $b=0.05$ times the mean inter-particle separation and then determined the spherical-overdensity (SO) mass centred on the centre-of-mass of the parent FoF group. A halo was defined as the sphere of radius $R_{200}$, which has a mean density of $200$ times the critical density, $\rho_{\rm crit}$. This corresponds to a halo mass $M_{200} = (4\pi/3)R_{200}^3(200\rho_{\rm crit})$.

Further, by analyzing all haloes individually, AHA13 found that the halo sample could be divided into various subsamples or ``types''. ``Type-1'' objects are virialized haloes, while ``type-2'' include haloes in late stages of formation; the latter do not show an isotropic density structure and instead contain larger scale caustics that are remnants of their formation. The remaining objects were haloes in early stages of formation, that have, e.g., just started collapsing along the third axis. 

\begin{figure}
\begin{center}
\includegraphics[width=0.45\textwidth]{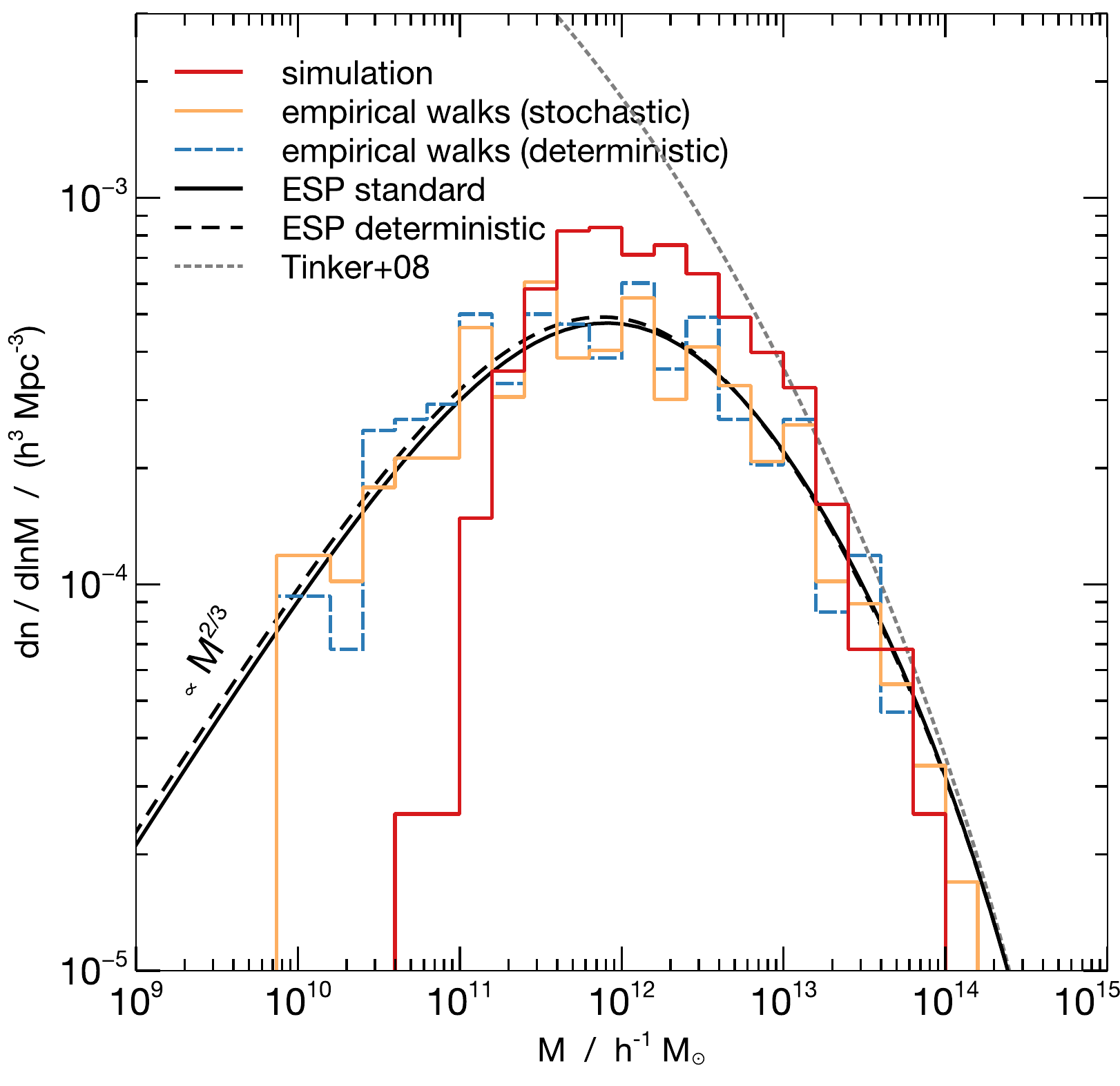}
\end{center}
\caption{Halo mass functions: the abundance of virialized objects in the simulation (red) as well as the mass functions of ``\peaks'' determined using empirical excursion set walk starting at peak locations from the same initial conditions as the simulation using the barrier in \eqn{simplebarrier} (orange; see Section~\ref{sec:empirical_walks_method} for details). The dashed blue histogram shows the result of the same algorithm but now using the deterministic barrier $B=\delc+0.5\sig_0$. The solid black line shows the ``standard'' analytic ESP calculation from \citet{psd13} which used a stochastic barrier adjusted to match CDM simulations. The dashed black line shows the ESP calculation using the deterministic barrier mentioned above (see Section~\ref{sec:analytical} for details). Although peaks theory correctly predicts a turn-over in the mass function at the correct scale, it gets the asymptotic behaviour below this scale wrong. For comparison, the dotted black line shows the result of using the \citet{Tinker08} fitting formula. 
}
\label{fig:mass-functions-empirical}
\end{figure}

In what follows, we only consider the ``type-1'' objects clearly identified as haloes. The red histogram in \fig{fig:mass-functions-empirical} is identical to the line labelled ``haloes'' in Figure~7 of AHA13 and shows the mass function of these ``type-1'' haloes, which has a sharp cut-off between $10^{11}$-$10^{12}\Ms$. (The other two histograms will be discussed in Section~\ref{sec:empiricalwalks} below.) It is important to note here that the classification was performed visually and we thus expect that, on an object-by-object basis, the distinction between type~1 and 2 is likely not perfectly robust. 

\subsection{Theoretical expectations}
\label{subsec:theoryexpect}
As discussed earlier, as far as late time structure formation is concerned, the only difference between the CDM and WDM cosmologies is the lack of {\em initial} small scale power in the latter. So one might expect that a physically motivated description of CDM structure should apply equally well to WDM -- at least at scales much larger than the free-streaming scale -- with the simple replacement of the CDM transfer function with the one in \eqn{Tk-WDM}. 
One can already anticipate that the standard hierarchical excursion set calculation would have difficulty in describing the low mass end of the WDM mass function where the effective logarithmic slope of the power spectrum  $n_{\rm eff}\equiv\der\ln P(k)/\der\ln k$ becomes steeper than $-3$. This is the boundary beyond which hierarchical prescriptions are known to fail, with predicted mass accretion rates becoming ill-defined \citep{lc93,lc94}. The ESP formalism, however, introduces a new ingredient into the picture -- the peaks constraint -- and since we know that it works well for CDM, we can ask what it predicts for the WDM case.

In what follows, we will frequently use integrals over the power spectrum of the filtered initial overdensity field $\del_R$ and its spatial derivatives, all linearly extrapolated to the present epoch:
\be
\sig_j^2(R) \equiv \int\der\ln k\,\Del(k)\,k^{2j}\tilde{W}_R(k)^2\,,
\label{sigma-j}
\ee
where $\Del(k)\equiv k^3P(k,z=0)/(2\pi^2)$ is the dimensionless matter power spectrum in linear theory and $\tilde{W}_R(k)$ is the Fourier transform of the smoothing filter, for which we will use a spherical TopHat $\tilde{W}_R(k) = 3\left(\sin kR - kR\,\cos kR \right)/(kR)^3$ in our numerical analysis and later also a Gaussian $\tilde{W}_R(k) = {\rm exp}{(-k^2R^2/2)}$ in our analytical modelling\footnote{All these integrals remain finite at all scales, including the unsmoothed limit $R\to0$, since the WDM free-streaming scale itself acts as a smoothing filter. In contrast, ultraviolet power in CDM causes $\sig_0(R)$ and $\sig_1(R)$ to diverge as $R\to0$, while the TopHat smoothed $\sig_2(R)$ always diverges, meaning that any analysis of small scale CDM peaks would be limited by effects at the spatial resolution limit of the simulation.}. The above definitions correspond to setting $\sig_0^2(R)=\avg{\del_R^2}$, $\sig_1^2(R)=\avg{(\nab\del_R)^2}$ and $\sig_2^2(R)=\avg{(\nabla^2\del_R)^2}$ and appear in peaks formalism calculations. Another quantity, which is relevant for excursion set models of the mass function, is the derivative of the smoothed density field with respect to smoothing scale, $\der\del_R/\der R$, which is in general different from the spatial derivatives of $\del_R$. A special case is that of Gaussian filtering, for which $\der\del_R/\der R = R\,\nabla^2\del_R$, a result which will be useful later in our analytical modelling.  For TopHat filtering, $\der\del_R/\der R$ and $\nabla^2\del_R$, although different, are strongly correlated \citep{psd13}.

To get an idea about the scale of the problem, consider that simply counting all peaks in the unsmoothed initial density field of the WDM simulation gives us 6713 objects, where ``unsmoothed'' refers to the density on a $512^3$ grid and a grid cell is labelled a peak if its density is higher than all its 26 neighbours (see Section~\ref{sec:haloprops} for more details). This grid size just about resolves the cutoff scale $\alpha$ (equation~\ref{alpha}) below which no initial fluctuations exist; we have verified that a $1024^3$ grid (the resolution at which the initial conditions of AHA13 were generated) leads to a consistent result (6822 peaks). This matches very well with the theoretical prediction for this number in the simulation volume $V_{\rm box}$ \citep[equation~4.11b of][BBKS from here on]{bbks86}:
\be
N_{\rm pk} = n_{\rm pk} V_{\rm box} = 3.12\times10^{-3}(\sig_2/\sig_1)^3 V_{\rm box} = 6608\,,
\notag
\ee
in which we evaluated $\sig_1$ and $\sig_2$ using \eqn{sigma-j} with $R=0$ and the transfer function \eqref{Tk-WDM}. Comparing this with the number of ``type-1'' objects in the simulation -- $1522$ -- we clearly see that not every individual peak forms a halo. This is fully expected within the analytical framework -- e.g., there is nothing special about a peak of height $\del = -10\sig_0$  -- and we will show later that the ESP calculation does lead to a number close to the measured number of haloes.

The ESP halo mass function can be written as \citep{ps12,psd13}
\be
\der n_{\rm ESP}/\der \ln m
 =  \Cal{N}_{\rm ESP}\,\left|\der\nu/\der\ln m\right|\,,
 \label{espansatz}
\ee
where $\nu\equiv\delc(z)/\sig_0(m)$ with $\delc(z)$ being the critical linear overdensity or ``barrier'' for spherical collapse in a $\Lambda$CDM background\footnote{The redshift dependence of $\delc(z)$ in a flat $\Lambda$CDM universe is slightly different from that in an Einstein-deSitter background \citep[see, e.g.,][]{ecf96}, and can be approximated by $\delc(z)D(z)/D(0) = \del_{\rm c,EdS}(1-0.0123\log_{10}(1+x^3))$, where $x\equiv(\Omega_{\rm m}^{-1}-1)^{1/3}/(1+z)$ and $\del_{\rm c,EdS}=1.686$ \citep{Henry2000}. In our case, requiring collapse at present epoch gives $\delc(z=0)=1.674$.}, and where $\Cal{N}_{\rm ESP}$ has the structure
\be
\Cal{N}_{\rm ESP} \sim \frac1{V_\ast}\times g_{\rm ESP}(\nu,\gam)\,.
\label{formofN}
\ee
Here $g_{\rm ESP}(\nu,\gam)$ is a dimensionless function of its arguments (details in Section \ref{sec:analytical}), and 
\gam\ and $V_{\ast}$ are spectral quantities that define the distribution of peaks (BBKS):
\be
\gam\equiv\sig_1^2/(\sig_0\sig_2) \quad;\quad V_\ast\equiv(6\pi)^{3/2}(\sig_1/\sig_2)^3\,.
\label{gam-Vst}
\ee
Whereas \gam\ is a dimensionless measure of the width of the power spectrum, $V_\ast$ sets the mean number density of all peaks on scale $R$ (equation~4.11b of BBKS). 

\begin{figure}
\begin{center}
\includegraphics[width=0.45\textwidth]{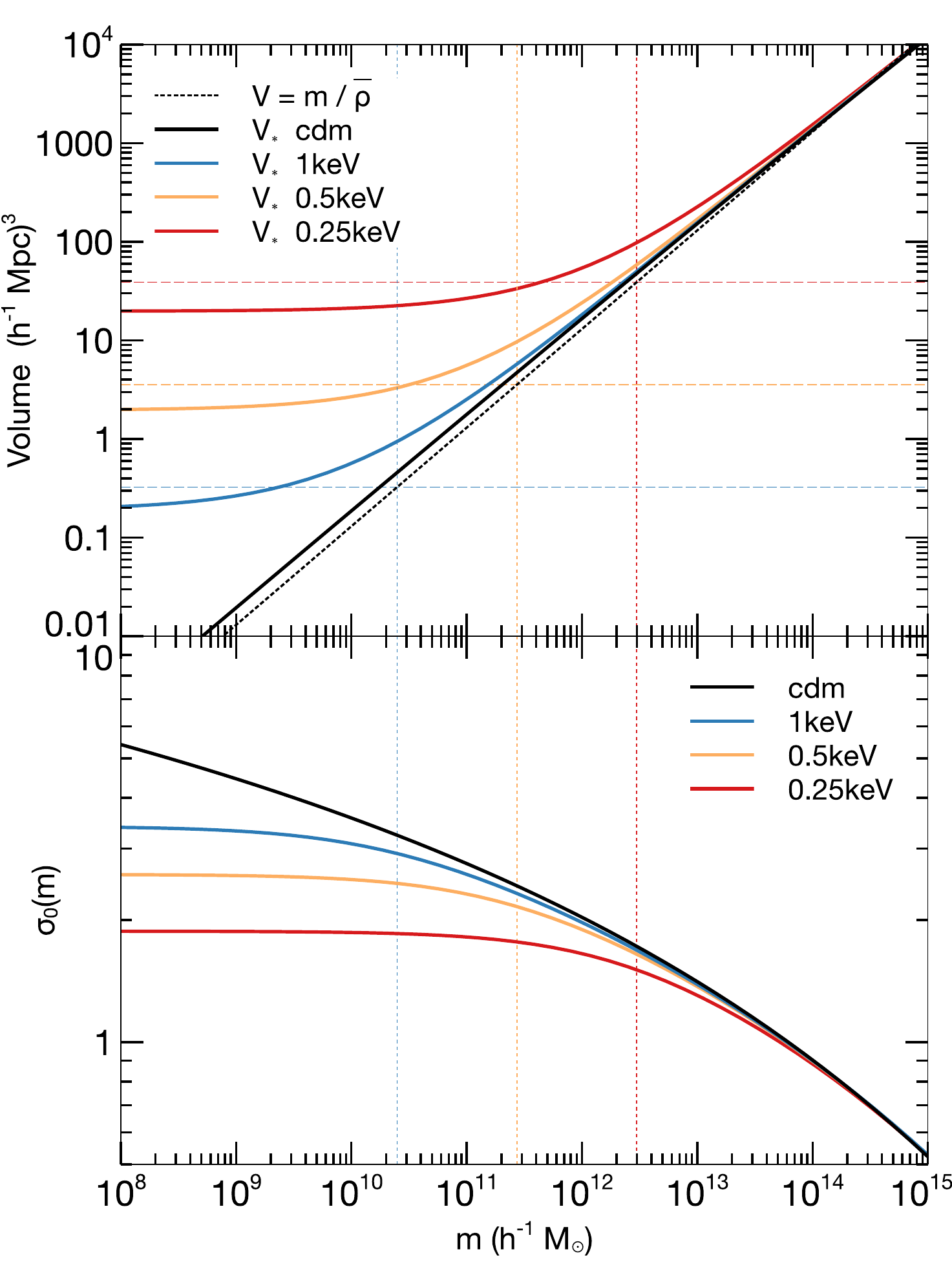}
\end{center}
\caption{(\emph{Top panel}:) Characteristic peak volume $V_\ast$ (solid curves) defined in \eqn{gam-Vst}, for CDM (black) and three choices of WDM particle mass $m_{\rm dm}$; from bottom to top: 1keV (blue), 0.5keV (yellow), 0.25keV (red). Dotted black line shows the Lagrangian volume $V=m/\bar\rho$. For CDM, $V_\ast$ closely tracks $V$. For WDM, as $m$ decreases, $V_\ast$ tracks $V$ until $m\sim M_{\rm hm}$ (vertical dotted lines) and then approaches a constant value close to $M_{\rm hm}/\bar\rho$ (horizontal dashed lines). All the $V_\ast$ curves used Gaussian filtering as described in Appendix~\ref{app:espStoch}. (\emph{Bottom panel}:) The relation between $\sig_0$ and mass $m$ with TopHat filtering for CDM (black) and when using \eqn{Tk-WDM} with the same particle masses (and colour-coding) as in the top panel. For WDM we clearly see a ``freezing-out'' of $\sig_0(m)$ at small masses.
}
\label{fig:vol}
\end{figure}

It is easy to see that the spectral integrals $\sig_j$, and consequently also $\nu$, \gam\ and $V_\ast$, will each approach a constant value for WDM as  $m\to0$. 
The behaviour of $\nu$ and $V_\ast$ in this respect is very different from that in CDM in the same limit, where $\nu,V_\ast\to0$ as $m\to0$. This reflects the fact that peaks can only form on scales large enough to be inhomogeneous; {\em reducing the smoothing scale cannot wipe out existing peaks for any power spectrum and, for a truncated spectrum, cannot introduce new peaks}. This can be seen in the top panel of\fig{fig:vol}: for each choice of $m_{\rm dm}$, $V_\ast$ at large $m$ is close to the CDM value (and hence approximately proportional to $V=m/\bar\rho$, where $\bar{\rho}$ is the comoving mean density), but deviates as $m$ approaches the half-mode mass $M_{\rm hm}$, finally approaching a constant value at $m\ll M_{\rm hm}$, this value being close to $M_{\rm hm}/\bar\rho$. This last aspect gives us an interesting physical interpretation of the half-mode mass as being essentially the same as the asymptotic peaks scale.

Finally, the Jacobian appearing in \eqn{espansatz} behaves, as $m\to0$ for WDM, like
\begin{align}
\left|\frac{\der\nu}{\der\ln m}\right| &= \frac{\nu^3 R}{3\delc(z)^2}\int\der\ln k\,\Del(k)W(kR)\left|\frac{\der W}{\der R}\right| \notag\\
&\to \frac{R^2\nu^3}{\#\delc(z)^2} \int\der\ln k\,\Del(k)k^2
\propto R^2 \propto m^{2/3}\,,
\label{jacobian-asymp}
\end{align}
where $\#=15$ for a TopHat filter and $3$ for a Gaussian filter,
so that the ESP mass function (equation~\ref{espansatz}) for WDM at small masses behaves like
\be
\der n_{\rm ESP}/\der\ln m~\sim|\der\nu/\der\ln m|~\sim~m^{2/3}\,.
\label{dndlnm-esp-asymp}
\ee
This power-law behaviour at small masses will be true for any single-barrier excursion set model of peaks in which the barrier depends on halo mass only through the spectral integrals; in particular, this behaviour is independent of details such as barrier shape, stochasticity, etc. 

The behaviour of $V_\ast$ discussed above shows that the turnover occurs at around $m\sim M_{\rm hm}$. The solid curve in \fig{fig:mass-functions-empirical} shows the ESP calculation of \citet{psd13} using the WDM transfer function \eqref{Tk-WDM}. The dashed curve shows the ESP result with a somewhat different, convenient choice of barrier for the random walks, which we will discuss in detail in Section \ref{subsec:stdESP} below. The main point is that both of these curves show the turnover and the asymptotic scaling of $\sim m^{2/3}$. As discussed above, the latter is a very robust prediction that cannot be easily altered by technical modifications within the framework of a single barrier. 

It is interesting to contrast this result with the corresponding one for the traditional excursion set approach, as this emphasizes the key role played by the behaviour of $V_\ast$. For traditional excursion sets \citep{bcek91,ms12}, one has
\be
\der n_{\rm trad}/\der \ln m
 = (\bar\rho/m)\, f(\nu)\,\left|\der\nu/\der\ln m\right|\,.
 \label{excsetansatz}
\ee
where $f(\nu)$ is a dimensionless function analogous to $g_{\rm ESP}(\nu,\gam)$ discussed earlier. More importantly, $V_\ast$ is replaced by the Lagrangian volume $V=m/\bar\rho$ of the halo, so that for WDM at small masses, this mass function behaves like
\be
\der n_{\rm trad}/\der\ln m~\sim~m^{-1}|\der\nu/\der\ln m|~\sim~m^{-1/3}\,.
\label{dndlnm-trad-asymp}
\ee
\emph{This asymptotic behaviour is unphysical because the hierarchical excursion set calculation should not predict objects at small masses where no hierarchical formation can occur in the absence of small scale power} (See the discussion above equation~\ref{sigma-j}). The dotted curve in \fig{fig:mass-functions-empirical} shows the result of using \eqn{Tk-WDM} to compute the relation $\sig_0(m)$ in the CDM fit provided by \citet{Tinker08}. For completeness, the bottom panel of \fig{fig:vol} compares the $\sig_0(m)$ relation for TopHat filtering when using \eqn{Tk-WDM} with the corresponding relation for CDM. For WDM we clearly see a ``freezing-out'' of $\sig_0(m)$ at small masses. The other spectral integrals also show similar behaviour.


\subsection{Theory vs. Simulation -- What could be going wrong?}
\label{sec:whatgoeswrong}

Although the physical requirement of being a density peak naturally accounts for a turnover in the mass function at the correct mass scale, the asymptotic scaling (very robustly) predicted by ESP is clearly wrong. There are several issues which could in principle affect this result:
\begin{description}
\item[{\bf Dynamics:}] A dramatic possibility is that, since small-mass haloes in WDM do not form hierarchically (e.g., at some point in time the first object forms, with no virialized progenitor), the peaks calculation might simply not be applicable. This would lead to the interesting question of just what it is that characterizes the locations and dynamics of the collapse of small mass objects. The cut-off scale in the initial spectrum could in principle allow for higher order catastrophes \citep[c.f.][]{Arnold1982} to become relevant, and these need not necessarily appear as peaks when filtered on the proto-halo scale. In CDM, the situation is quite different in this respect, since fluctuations persist down to very small scales and so every proto-halo has a progenitor at a smaller scale.

\item[{\bf Barrier shape:}] It has been argued that the collapse barrier appropriate for WDM haloes is very different from the corresponding CDM one due to thermal effects in WDM, and that this can introduce a sharp cut-off in the mass function \citep{b+13}. However, since WDM simulations see a cut-off despite ignoring thermal effects \citep[e.g., AHA13;][]{ssr13}, the origin of the cut-off must be rooted in the suppression of \emph{initial} small scale power, and must therefore be a generic feature of cold collisionless dynamics in such conditions. 
The failure of excursion set (peaks) models to reproduce the correct mass function hence indicates quite clearly that these models are still not accounting for some important physical processes. One of the primary goals of this work is to investigate the cause of this behaviour. 

\item[{\bf Patch shape:}] Traditional excursion sets, as well as the ESP calculation of \citet{psd13}, use spherical filters when assigning masses to objects, and it could be that asphericity of the Lagrangian patches affects the mass assignment significantly at small masses. E.g., recently \citet*{dts13} have demonstrated using CDM simulations that accounting for halo asphericity using an ellipsoidal halo finder can lead to small increases in mass for low mass haloes \citep[see also][]{lp11-collapse}.

\item[{\bf Stochasticity:}] Regardless of the importance of thermal effects, the specific details of the barrier, e.g., those related to stochasticity in the barrier height, are in fact somewhat uncertain (even in the CDM case). The ESP calculation for CDM is self-consistent but not fully predictive, and needs some inputs from simulations \citep{psd13}. In particular, the barrier used in that calculation was adjusted to match measurements by \citet{Robertson2009} of proto-halo overdensity in CDM simulations, and the same results might not apply in the case of WDM.

\item[{\bf Peak-in-peak:}] Another possible source of error is that the ESP framework treats the peak-in-peak problem approximately, by introducing the effects of the peaks constraint as an extra weight in the mass function, rather than by explicitly accounting for spatial correlations between walks centred at different locations in space \citep[see, e.g.,][]{sb02}, and this approximation needs testing.
\end{description} 

\noindent We address these issues in the next two Sections by exploring the properties of the initial conditions of the simulation in greater detail\footnote{A further role might be played by assembly bias, i.e., the dependence of halo formation histories on scales substantially larger than the Lagrangian patch. Assembly bias is typically seen as a suppression of late-time growth for low-significance haloes \citep[c.f., e.g.,][]{st04,Gao2005,Desjacques08,Hahn2009,Fakhouri2010}. The impact of large-scale tidal fields on the collapse of scales around the half-mode scale, where structure formation is not hierarchical, has (to our knowledge) not been studied yet. This aspect would clearly be of interest in future work. }.


\section{Lagrangian Properties of Haloes}
\label{sec:haloprops}

In this Section, we turn to the initial conditions of the simulation and perform an in-depth study of the Lagrangian properties of regions that will eventually form haloes; we call such regions proto-haloes and give a precise definition below. 
Several authors have performed such studies in CDM simulations \citep[e.g.,][]{White1996,bm96,smt01,pdh02,Robertson2009,lp11-collapse,lp11,elp12,dts13}. 
To our knowledge, the current work is the first to extend these studies to the case of WDM, and is interesting for the reasons discussed in Section~\ref{sec:massfunc}.

An advantage of using a WDM model with $m_{\rm dm}=0.25$keV is that the number of objects is reasonably small. A disadvantage is that the half-mode mass is close to being unit-significance, $\nu(M_{\rm hm},z=0)\simeq0.9$, which does not allow us to explore low-significance objects with sufficient statistical precision. This could also potentially confuse non-linear assembly-bias-like effects with the peculiarities of halo formation at and below the half-mode mass scale. Nevertheless, this simulation provides us with an invaluable testing ground for several ideas in the peaks framework.

We focus on the overdensity of the proto-halo patch (which is indicative of the collapse threshold), its curvature, velocity shear (ellipticity and prolateness) and moment of inertia. We will demonstrate two important features of the proto-halo patches; (a) that they are all consistent with forming at initial density peaks and (b) their overdensities are strongly correlated with their ellipticities but not their prolateness.

\subsection{Lagrangian density and shear fields}
\label{sec:lagrange_density}
The initial conditions code {\sc Music} allows us to output the density field that was used to generate the simulation initial conditions as three-dimensional grid data. We used this function to re-generate the density field directly on a $512^3$ mesh with the same Fourier modes as the original simulation. We refer to this as the unsmoothed field \del. Using the particle IDs that encode the three dimensional Lagrangian coordinate $\mathbf{q}$ on the unperturbed initial particle lattice (i.e. before applying the Zel'dovich approximation), we can directly evaluate the density at $\mathbf{q}$ without interpolating the perturbed particle position back on a grid. We linearly scale the density field to $z=0$.

Using the unsmoothed density field on a mesh, we compute various derived fields using the fast Fourier transform (FFT). We compute the gradient and the Hessian of the density field,
\begin{equation}
\nab\delta = \mathcal{F}^{-1}\left\{ i\mathbf{k}\tilde{\delta}(\mathbf{k})\right\},\quad \partial_{ij}\delta = \mathcal{F}^{-1}\left\{ -k_i k_j \tilde{\delta}(\mathbf{k})\right\},
\end{equation}
where the tilde indicates the Fourier transformed field. Additionally, we compute the velocity potential as well as its Hessian (the so-called tidal tensor which reflects the velocity shear),
\begin{equation}
\psi = \mathcal{F}^{-1}\left\{ -k^{-2} \tilde{\delta}(\mathbf{k})\right\}, \quad \partial_{ij}\psi = \mathcal{F}^{-1}\left\{ \frac{k_i k_j}{k^{2}} \tilde{\delta}(\mathbf{k}),\right\},
\end{equation}
so that the velocity field is $\mathbf{u}\propto-\boldsymbol{\nabla}\psi$. When computing filtered fields, we replace $\tilde{\delta}(\mathbf{k})$ with $\tilde{\delta}_R(\mathbf{k})=\tilde{\delta}(\mathbf{k})\tilde{W}_R(k)$. 

We define the ordered eigenvalues of $\partial_{ij}\delta$ as $\zeta_1\le\zeta_2\le\zeta_3$ and those of the velocity shear $\p_{ij}\psi$ as $\lambda_1\le\lambda_2\le\lambda_3$. The normalised negative trace of the density Hessian gives us the dimensionless peak curvature 
\be
x\equiv-(\zeta_1+\zeta_2+\zeta_3)/\sig_2\,,
\label{x-def}
\ee
while the trace of the velocity shear gives back the density
\be
\delta={\rm Tr}\,\partial_{ij}\psi = \lambda_1+\lambda_2+\lambda_3 \propto -{\rm div}\,\mathbf{u}\,.
\label{tidal-trace}
\ee
Peaks in $\delta$ are thus equivalent to regions of maximum convergence in the Lagrangian flow. We will also need the ellipticity $e_v$ and prolateness $p_v$ associated with the tidal tensor:
\begin{align}
e_v\del &\equiv \left(\lambda_3-\lambda_1\right)/2 \equiv Y\,,
\label{ev-Y-def}\\
p_v\del &\equiv \left(\lambda_3-2\lambda_2+\lambda_1\right)/2 \equiv Z\,,
\label{pv-Z-def}
\end{align}
where we have defined $Y$ and $Z$ to be the corresponding un-normalised quantities which will be useful below. Similarly, we can define
\begin{align}
e_{\rm pk}x &\equiv-(\zeta_1-\zeta_3)/(2\sig_2) \equiv y\,,
\label{epk-y-def}\\
p_{\rm pk}x &\equiv -(\zeta_1-2\zeta_2+\zeta_3)/(2\sig_2) \equiv z\,,
\label{ppk-z-def}
\end{align}
so that $e_{\rm pk}$ and $p_{\rm pk}$ describe the shape of the peak (BBKS).


\subsection{Proto-haloes and their properties}
For each halo, we recorded the particle IDs and recovered their respective Lagrange coordinates $\mathbf{q}$.
We call the set of $N_k$ Lagrangian particles comprising halo $k$ its Lagrangian patch or proto-halo, denoted $L_k\equiv\left\{\mathbf{q}_i\,\left| \,i=1\dots N_k \right. \right\}$.

\begin{figure}
\begin{center}
\includegraphics[width=0.275\textwidth]{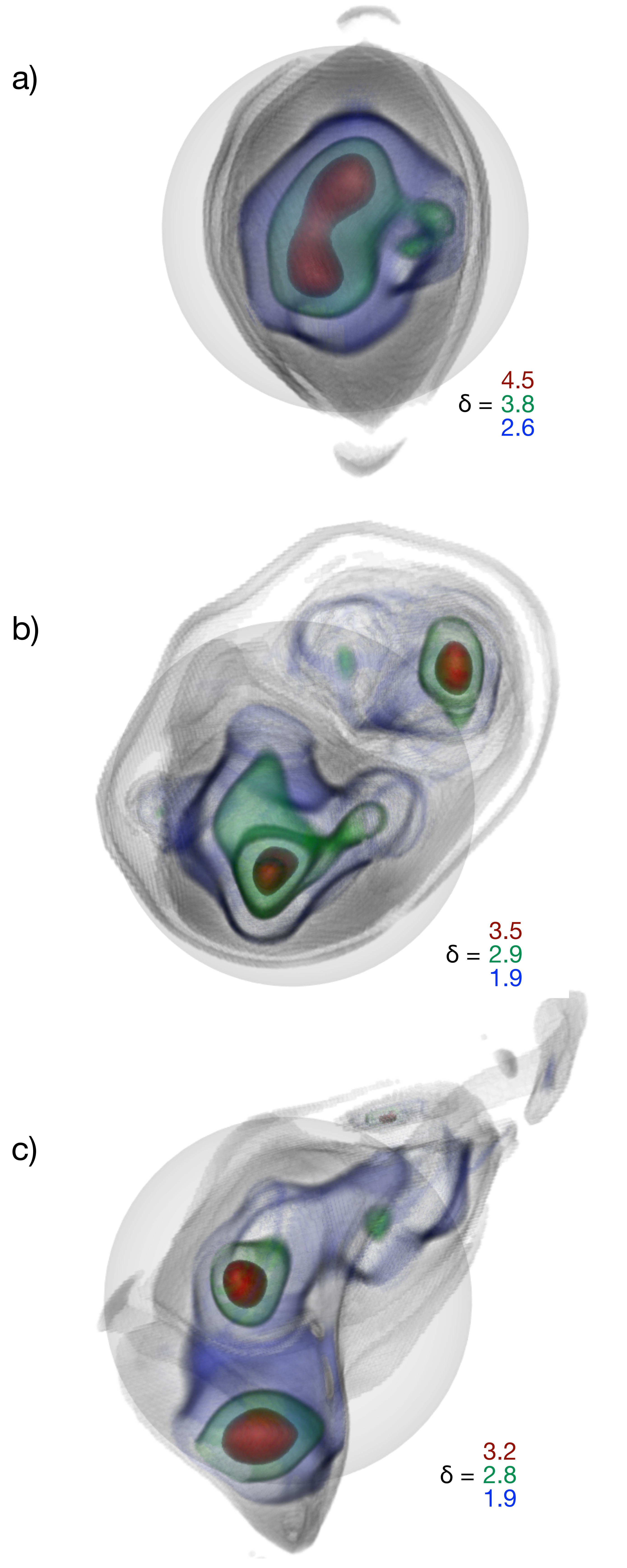}
\end{center}
\caption{Volume renderings of the (unsmoothed) density field in the proto-halo patches for three haloes of mass $10^{13}\Ms$. Each panel shows a light grey sphere centred on the proto-halo centre, containing the same mass. The darker shaded gray volume indicates the actual proto halo patch. Coloured contours indicate isodensity regions at the values given underneath each image. (\emph{Top panel}:) A good example of an ellipsoidal proto-halo patch, with clearly visible disconnected outer regions. This proto-halo was also assigned a matching \peak\ by the algorithm described in Section~\ref{sec:empiricalwalks}. (\emph{Middle \& bottom panels}:) Two examples of proto-haloes with evidence for substantial substructure and mixing due to large scale interactions. Our algorithm did not find any matching \peak\ for these two objects.
}
\label{fig:lagrange_examples}
\end{figure}

We compute the patch average of a Lagrangian field $f(\mathbf{q})$ by evaluating
\begin{equation}
\left< f\right>^{\rm (p)}_k = \frac{1}{N_k}\sum_{\mathbf{q}_i\in L_k} f( \mathbf{q}_i ).
\end{equation}
and the spherical average by first determining the Lagrange radius $R_{\rm L}=(3m/4\pi \bar{\rho})^{1/3}$, where $m$ is the halo mass, and then evaluating
\begin{equation}
\left< f\right>^{\rm (s)}_k = (f\otimes W_{R_{\rm L}}) (\mathbf{q}_\textrm{med}).
\end{equation}
Here $W_{R_{\rm L}}$ is the TopHat filter at scale $R_{\rm L}$ and $\mathbf{q}_\textrm{med}$ is the median Lagrange coordinate of the Lagrangian patch, where the median is taken of each separate Cartesian component. Using the median instead of the mean coordinate reduces the influence of outliers in the Lagrangian patch. 

Figure~\ref{fig:lagrange_examples} shows three examples of proto-haloes with masses $\sim10^{13}\Ms$. The top panel shows a well behaved proto-halo. We notice two disconnected shells surrounding the connected interior of this patch. This is a beautiful example of the mapping between Lagrangian and Eulerian space. The gaps between the shells appear because the outer caustics of the halo are not inside the virial radius and are thus cut off. The two shells correspond to material on first and second infall. The other two examples show evidence of mixing due to large scale interactions.

In addition to the ellipticity and prolateness associated with the density Hessian, we can also characterize the shape of the Lagrangian patch $L_k$ through  the dimensionless reduced moment of inertia tensor
\begin{equation}
{\rm I}_{ij} = \sum_{\mathbf{q}\in L_k}\left( \mathbf{q}^2 \delta_{ij} -  q_i q_j \right)/\mathbf{q}^2\,,
\end{equation}
which we define to be centred on the centre-of-mass of the object (rather than its median location), since this minimises its values. The eigenvalues $\iota_1\leq\iota_2\leq\iota_3$ of ${\rm I}_{ij}$ give the corresponding axes of the homogeneous ellipsoid $a\leq b\leq c$:
\begin{align}
a & =  \sqrt{5/2N_k} \left( \iota_1+\iota_2-\iota_3\right)\,,\notag\\
b & =  \sqrt{5/2N_k} \left( \iota_1-\iota_2+\iota_3\right)\,,\notag\\
c & =  \sqrt{5/2N_k} \left( -\iota_1+\iota_2+\iota_3\right)\,,
\end{align}
and the sphericity
\begin{equation}
S \equiv a/c\,.
\label{sphericity-def}
\end{equation}
%


\subsection{Haloes form at peaks}

We start by verifying statistically that haloes in cosmologies with truncated small-scale power do indeed form from peaks. Being a peak requires the overdensity field $\delta$ to be locally extremal on the scale of the proto-halo, i.e.
\begin{equation}
\left<\boldsymbol{\nabla}\delta\right>=\mathbf{0}\quad\textrm{and}\quad \left<\zeta_i\right><0,\,i=1,2,3,
\label{pk-constraint}
\end{equation}
We find that all proto-halo patches have $\left<\zeta_i\right>^{\rm (s)}<0$ and $\left<\zeta_i\right>^{\rm (p)}<0$ and thus the total peak curvature $x$ is positive in both cases. Note that to define averaged eigenvalues of a tensor, we diagonalise after computing the average of the tensor.

\begin{figure}
\begin{center}
\includegraphics[width=0.4\textwidth]{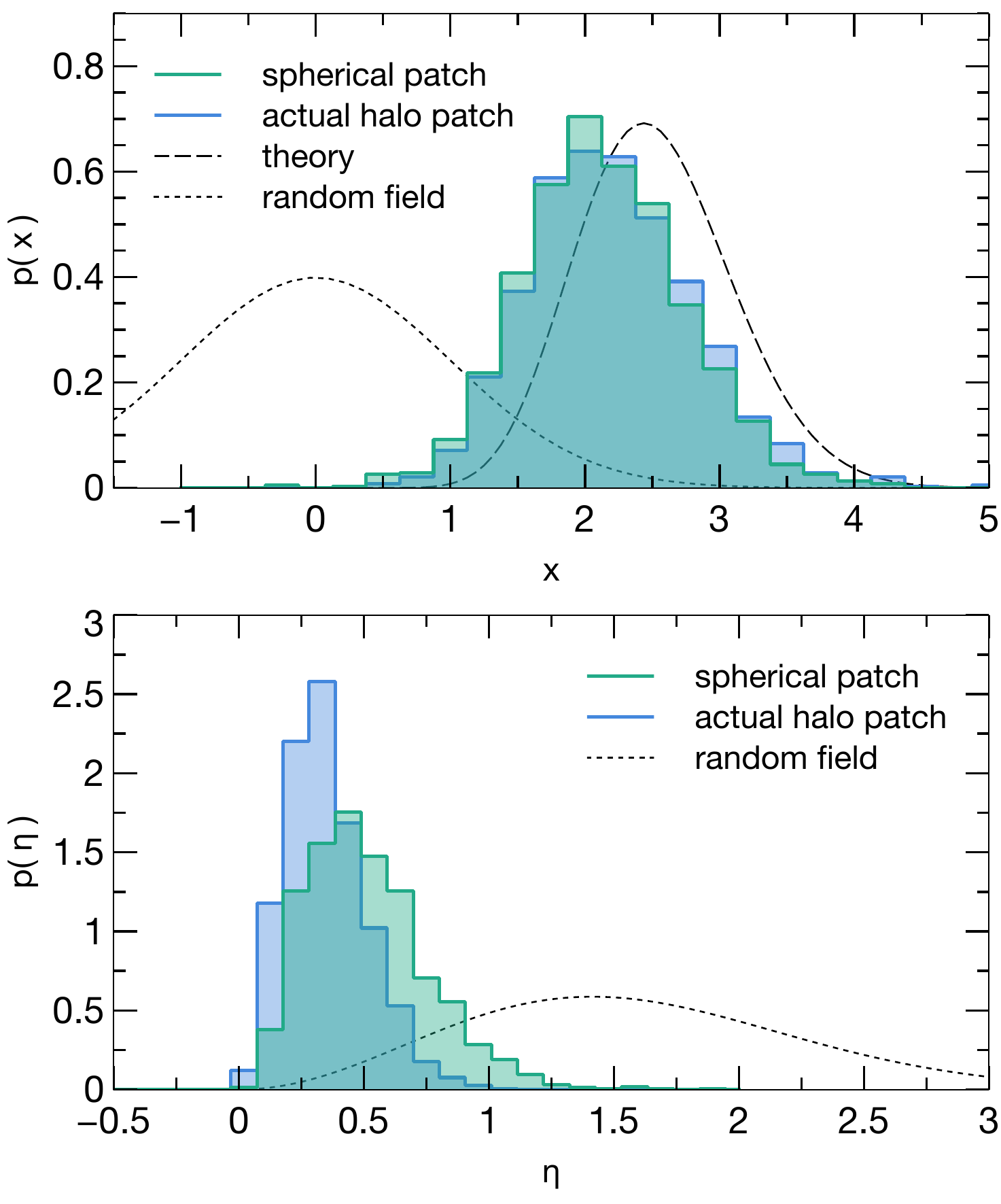}
\end{center}
\caption{(\emph{Top panel}:) The distribution of peak curvatures $x$ (equation~\ref{x-def}) at proto-halo locations, averaged over the Lagrangian patch (blue histogram) and using a spherical aperture of the same mass (green histogram). The dashed curve shows the theoretically expected distribution (equation~\ref{px-esp}) while the dotted curve shows the distribution at random locations (a Gaussian with zero mean and unit variance). (\emph{Bottom panel}:) The distribution of density gradient $\eta\equiv\sqrt{3}|\nab\del|/\sig_1$ at proto-halo locations, averaged over the Lagrangian patch (blue histogram) and using a spherical aperture of the same mass (green histogram). The dotted line is the distribution at random locations; in this case $\eta^2$ is distributed as Chi-squared with 3 degrees of freedom. These plots show that all haloes in our sample are consistent with having formed near initial density peaks.}\label{fig:peak_prop_distribution}
\end{figure}

In Figure~\ref{fig:peak_prop_distribution}, we show the distribution of the total curvature $x$ (equation~\ref{x-def}; top panel) and the magnitude of the gradient $\eta\equiv\sqrt{3}|\nab\del|/\sig_1$ (bottom panel) averaged over the halo patches. For computing $\sig_1$ and $\sig_2$ which make these quantities dimensionless, we used a TopHat filter at the Lagrangian scale $R_{\rm L}$ of each object. 

The distribution of $x$ for a Gaussian random field would be a Gaussian with zero mean and unit variance (dotted black curve in the top panel). The measured distribution on the other hand has only positive values as mentioned above, and its shape is very similar to the analytical prediction using ESP with a deterministic barrier (dashed black curve, see Section~\ref{sec:analytical}, equation~\ref{px-esp}), although the measured mean value for $x$ is lower than the predicted mean by about $0.4$.

The distribution of $\eta$ for a Gaussian random field would be $p(\eta)=\sqrt{2/\pi}\eta^2{\rm e}^{-\eta^2/2}$ (because in this case $\eta^2$ is Chi-squared distributed with $3$ degrees of freedom.) This is shown as the dotted black curve in the bottom panel; the measured values clearly populate the low tail of this distribution. (Ideally all the values would be zero.) We also see that the patch-averaged values of $\eta$ have a significantly lower scatter than the spherically averaged ones. This is not surprising since the requirement $\eta=0$ is quite unstable to choices of filtering, and the spherical filter is known to introduce an additional randomisation as compared with the actual Lagrangian patch \citep[BBKS;][]{dts13}. 

We therefore conclude that all haloes in our sample are consistent with having formed near initial density peaks.
In principle, we should also have explicitly checked for the presence of local density maxima at or near the proto-halo locations, e.g., along the lines discussed by \cite{lp11}. This, however, would involve making a specific choice regarding the smoothing scale. We defer such a calculation to Section~\ref{sec:empirical_walks_method}, where we implement an algorithm that makes this choice while simultaneously centering the smoothing filter at locations that are most likely to collapse according to the excursion set formalism.

\subsection{Overdensity of Lagrangian patches}

\begin{figure*}
\begin{center}
\includegraphics[width=0.95\textwidth]{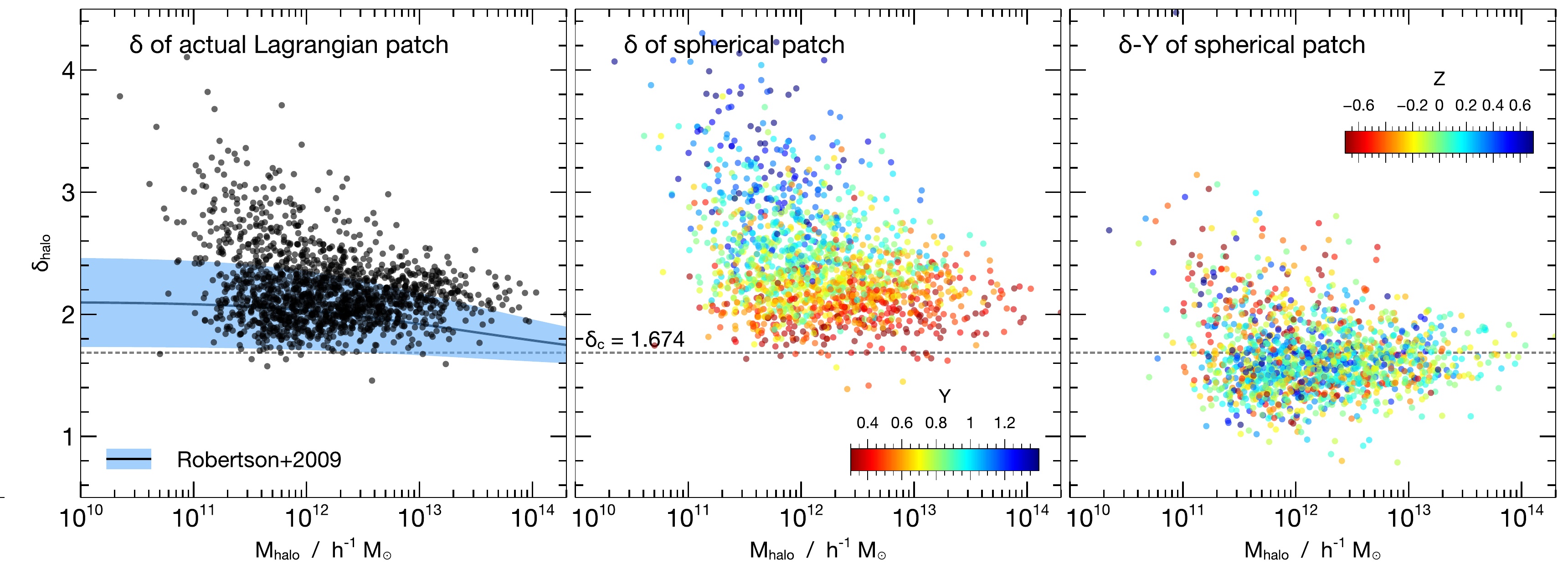}
\end{center}
\caption{Patch overdensity as a function of proto-halo mass. (\emph{Left panel}:) The overdensity \del\ (equation~\ref{tidal-trace}) at proto-halo locations, averaged over the Lagrangian patch (points). The thick blue line and the blue band show, respectively, the mean and standard deviation of the spherically averaged proto-halo overdensities measured by \citet{Robertson2009} in CDM simulations, extrapolated to our WDM initial power spectrum (i.e., we used our WDM transfer function in the fit in their equation D2 with the $\Delta=200$ parameters from their Table 1). (\emph{Middle panel}:) Proto-halo overdensity \del\ averaged over spherical apertures of the same mass, coloured by the spherically averaged ellipticity $Y=e_{\rm v}\del$ (equation~\ref{ev-Y-def}). We see a strong trend of \del\ with $Y$, as expected from ellipsoidal collapse arguments \citep{smt01}. (\emph{Right panel}:) The quantity $\del-Y$, coloured by the spherically averaged prolateness $Z=p_{\rm v}\del$ (equation~\ref{pv-Z-def}). The residual scatter shows no trend with $Z$, which is at odds with the predictions of ellipsoidal collapse models.
}
\label{fig:peakdensities}
\end{figure*}

Figure~\ref{fig:peakdensities} shows the patch-averaged (left panel) and spherically averaged (middle panel) overdensities of the proto-haloes as a function of their mass. We find that the spherical overdensities are strongly correlated with the corresponding spherically averaged values of $Y$ (equation~\ref{ev-Y-def}). This is evident in the middle panel where we have coloured the points using $\avg{Y}^{\rm (s)}$. (The patch-averaged overdensities show a similar strong correlation with the patch-averaged $Y$; we omitted the colouring in the left panel for clarity.) 

The right panel of the Figure shows the difference $\avg{\del}^{\rm (s)}-\avg{Y}^{\rm (s)}$ coloured by prolateness $\avg{Z}^{\rm (s)}$ (equation~\ref{pv-Z-def}). We see that the scatter in this difference is significantly smaller than that in $\avg{\del}^{\rm (s)}$, and its distribution is curiously similar to that of $\avg{\del}^{\rm (p)}$ with a mean close to the standard spherical collapse value \delc\ (horizontal dashed line in all panels). More importantly, we have found no correlation of $\avg{\del}^{\rm (s)}$ with prolateness $\avg{Z}^{\rm (s)}$. In other words, the spherical overdensities of the proto-haloes are well approximated by the relation 
\be
\avg{\del}^{\rm s} = \delc + \avg{Y}^{\rm s}\,,
\label{empiricalbarrier}
\ee
with a residual scatter that does not correlate with prolateness. This is consistent with the CDM results of \citet{lp11-collapse}.
We have also checked that the overdensity does not correlate with the other shape parameters $y$ and $z$ defined in \eqns{epk-y-def} and~\eqref{ppk-z-def}.

\citet{Robertson2009} performed similar spherically averaged measurements in the initial conditions of the CDM simulations presented by \citet{Tinker08}. The left panel of Figure~\ref{fig:peakdensities} shows the mean and standard deviation of the distribution of overdensities reported by \citet{Robertson2009}, but using the WDM transfer function. We see that their spherically averaged measurements, extrapolated to WDM, are in reasonable agreement with our patch-averaged overdensities. Our spherical overdensities, on the other hand, have a higher mean and scatter than theirs \citep[see also][who found similar results in their CDM simulations]{elp12}. There is no clear reason for this discrepancy. 

These results for the spherically averaged overdensity and shear ellipticity will form the basis of the empirical walks that we describe in the next Section. In general, however, we note that our measurements of spherically averaged quantities tend to have larger scatter than the corresponding patch-averaged ones. For completeness, in Appendix~\ref{app:peakshapes} we also show the distributions of ellipticity and prolateness defined using the tidal tensor and the Hessian of the density.

In summary, the results of this section show us that haloes form at peaks and have Lagrangian (spherically averaged) overdensities that are consistent with \eqn{empiricalbarrier}, with a residual scatter that is uncorrelated with other properties such as shear prolateness or peak shapes. One aspect we have not explored here is the relative (mis-)alignment between the velocity shear, density Hessian and moment of inertia tensors, which can be an important ingredient in any recipe for predicting collapse-time based on dynamical arguments. This is especially interesting given previous results from CDM simulations \citep{pdh02,dts13} which suggest that, contrary to expectations based on Gaussian statistics \citep[e.g.,][]{vdwb96}, the direction of maximum initial compression is on average well-aligned with the longest geometrical axis of the proto-halo, rather than the shortest. We will return to an analysis of tensor alignments and their dynamical consequences in future work.


\section{Empirical Excursion Set Peak Walks}
\label{sec:empiricalwalks}
The most serious issue raised in Section~\ref{sec:whatgoeswrong} was whether or not the excursion set formalism can capture {\em at all} the formation of haloes in WDM, which does not proceed hierarchically below the half-mode mass scale. In excursion set language, this amounts to asking whether or not the relation 
\be
B = \delc + Y
\label{simplebarrier}
\ee
actually works as a barrier for random walks of the density centred on peaks, and whether the resulting objects predicted to collapse from peaks at specific locations with a certain mass bear any relation to the haloes found in the simulation.

To test this, in this Section, we explicitly perform such random walks in the actual initial density field that was used for the numerical simulation and identify spherical peak-patches which are predicted to form haloes using this barrier. We will refer to these objects as \peaks\ below. 

Note that the relation~\eqref{simplebarrier} is similar to that predicted by the dynamics of a collapsing homogenous ellipsoid, which is well-approximated by \citep{smt01}
\be
B_{\rm ec} = \delc\left(1+\beta_{\rm ec}\left[5\left(Y^2\pm Z^2\right)/\delc^2\right]^{\gam_{\rm ec}}\right)\,,
\label{SMTbarrier}
\ee
where $\beta_{\rm ec}=0.47$ and $\gam_{\rm ec}=0.615$ and the minus (plus) sign is to be used when $Z$ is positive (negative). The most important difference is the absence of the prolateness in \eqn{simplebarrier}. We will return to this issue below.


\subsection{Methodology}
\label{sec:empirical_walks_method}
Our algorithm is essentially a more accurate version of what the analytical ESP calculation tries to achieve. We note that it is less sophisticated than the original peak-patch algorithm implemented by \citet{bm96}, since we are not interested in the final locations, profiles and velocity dispersions of the haloes, but only in their mass.

We consider a hierarchy of $N_s=100$ smoothing scales $R_i$ logarithmically spaced in the TopHat spherical mass contained in $R_i$ between $M_0=10^{10}\Ms$ and $M_{N_s}=10^{15}\Ms$.
We then proceed as follows, starting with the largest smoothing scale:

\begin{enumerate}[1.]
\item We determine the coordinates $\mathbf{x}_k$ of all peaks in $\delta_{R_i}$. Being a peak requires that $\delta_{R_i}$ is larger at $\mathbf{x}_k$ than in all 26 surrounding cells.
\item We discard all peaks for which $\delta_{R_i}$ is below the barrier, i.e. where $\delta_{R_i}(\mathbf{x}_k)< B(\mathbf{x}_k;R_{i})$; $B$ is given by \eqn{simplebarrier}.
\item Additionally, we discard all peaks that are within the Lagrangian radius of a peak that has been identified before. This explicitly solves the cloud-in-cloud problem. 
\item Finally, we also discard all peaks where the density was above threshold on a larger scale, i.e. where $\delta_{R_{i+1}}(\mathbf{x}_k)>B(\mathbf{x}_k;R_{i})$. This step improves numerical stability but is otherwise redundant.
\item We proceed to the next smaller scale $i\rightarrow i-1$ and start over at step 1.
\end{enumerate}

\noindent 
A small fraction of objects have partially overlapping Lagrangian volumes. We flag the smaller of such pairs as ``sub-peaks'' and, for the current analysis, do not include them in the sample of proto-haloes. (These form about $10\%$ of the total sample.) At the end, we arrive at a catalogue of \peaks\ whose Lagrangian properties we can analyse in exactly the same way as for the actual proto-halo patches.

\subsection{The mass function of \peaks}
\label{subsec:empirical-massfunc}
The solid orange histogram in Figure~\ref{fig:mass-functions-empirical} corresponds to the mass function of \peaks\ obtained using the algorithm described above. The dashed blue histogram shows the result of the same algorithm, but now using a \emph{deterministic} barrier \eqn{simpledetbar} which, as we argue in the next section, is a useful approximation to \eqn{simplebarrier}. Indeed, we see that these two histograms agree quite well, indicating that stochasticity in the barrier arising from statistical fluctuations in the initial conditions does not lead to dramatic effects in the mass function. 

\begin{figure}
\begin{center}
\includegraphics[width=0.4\textwidth]{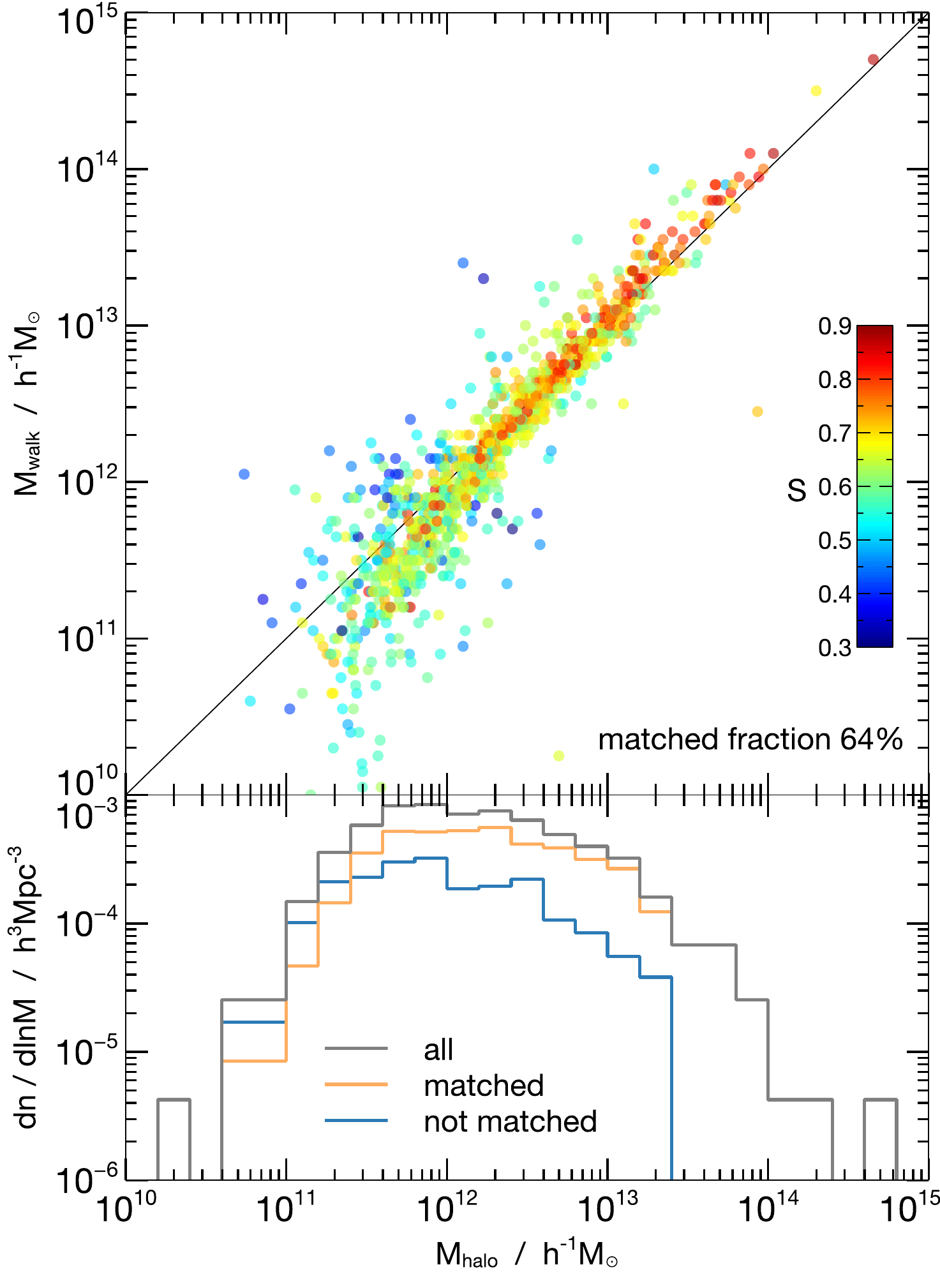}
\end{center}
\caption{(\emph{Top panel}:) The masses $M_{\rm halo}$ of proto-haloes that could be matched to \peaks, compared to the corresponding \peak\ mass $M_{\rm walk}$, coloured by the spherically averaged proto-halo sphericity $S$ (equation~\ref{sphericity-def}). Low mass proto-haloes are clearly more aspherical than high mass ones, and the scatter in mass assignment also increases significantly for the more aspherical objects. At a given halo mass, however, the average mass mismatch does not correlate strongly with $S$. (\emph{Bottom panel}:) The orange (blue) histogram shows the mass function of proto-haloes that could (not) be matched to \peaks. The gray histogram shows the total halo mass function (same as the red histogram in Figure~\ref{fig:mass-functions-empirical}). The matched fraction is quite large at high masses and falls significantly at low masses.
}
\label{fig:matching_mass_halo2walk}
\end{figure}

\begin{figure}
\begin{center}
\includegraphics[width=0.4\textwidth]{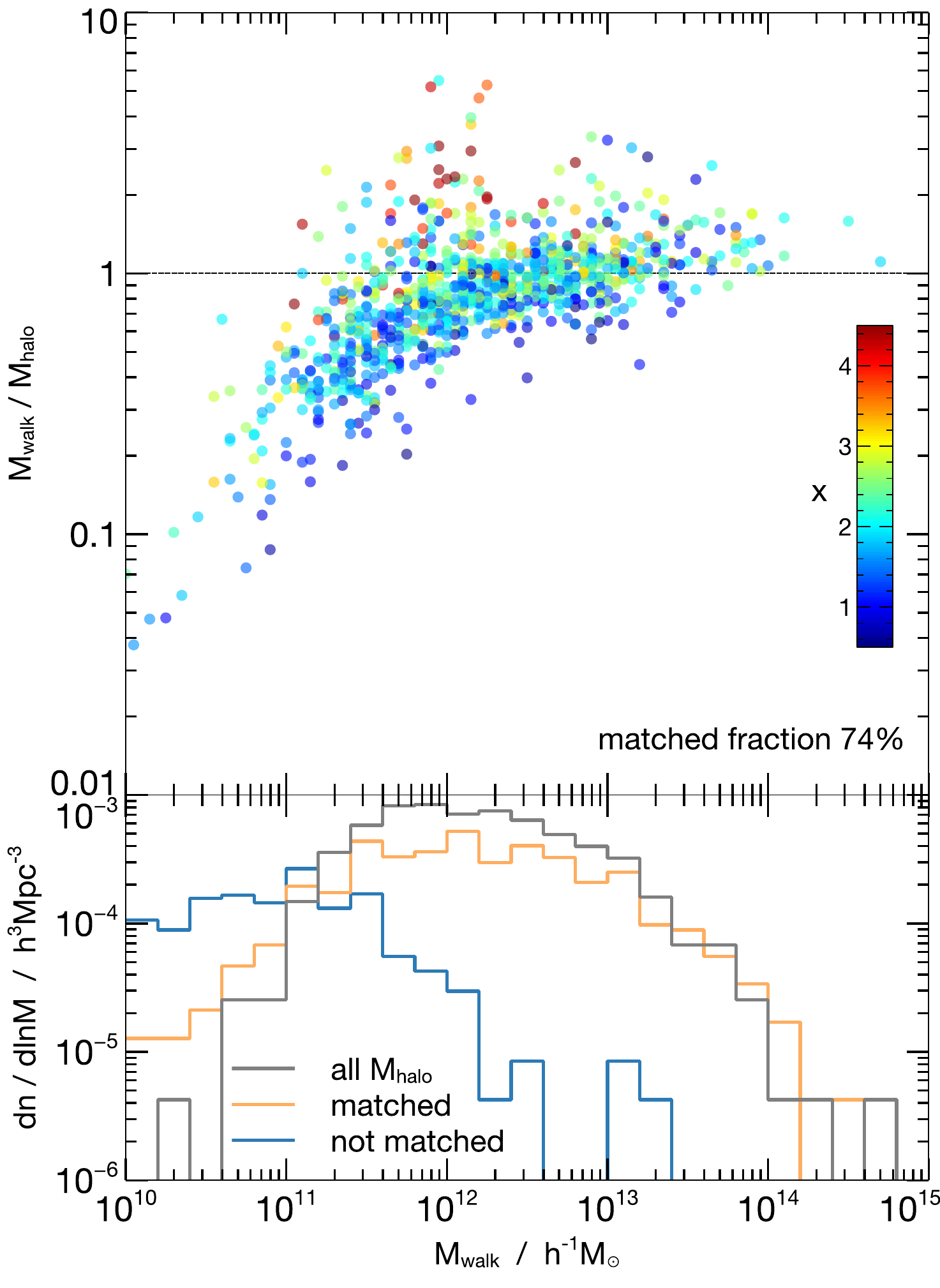}
\end{center}
\caption{(\emph{Top panel}:) The masses $M_{\rm walks}$ of \peaks\ that could be matched to proto-haloes, compared to the corresponding proto-halo mass $M_{\rm halo}$, coloured by the spherically averaged proto-halo curvature $x$ (equation~\ref{x-def}). 
We see a weak trend of the mass mismatch with curvature: shallower proto-haloes tend to have larger mismatches between $M_{\rm halo}$ and $M_{\rm walk}$. (\emph{Bottom panel}:) The orange (blue) histogram shows the mass function of \peaks\ that could (not) be matched to proto-haloes. The gray histogram shows the total halo mass function (same as the red histogram in Figure~\ref{fig:mass-functions-empirical}). Most of the unmatched \peaks\ were assigned dramatically lower masses than any proto-halo in the sample. The sum of the orange and blue histograms is consistent with analytical ESP predictions (compare the smooth black curves and solid orange histogram in Figure~\ref{fig:mass-functions-empirical}).
}
\label{fig:matching_mass_walk2halo}
\end{figure}

The most important feature of the orange histograms is that they show a low mass tail consistent with $\der n/\der\ln m \propto m^{2/3}$ as discussed earlier. In fact, the histograms are well described by the WDM version of the ESP calculation (solid black) of \citet{psd13} who used a stochastic barrier adjusted to match CDM simulations, as well as a similar ESP calculation with the deterministic barrier \eqref{simpledetbar} which we discuss below. 

The overall number of \peaks\ identified by our algorithm ($1261$ when using equation~\ref{simplebarrier} and $1335$ when using equation~\ref{simpledetbar}) is reasonably close to the total number of proto-haloes, which is $1522$. The lower numbers of \peaks\ could partially be because we stop our algorithm at the lower mass limit of $10^{10}\Ms$. For comparison, integrating the ESP prediction using \eqn{simpledetbar} (dashed line in Figure~\ref{fig:mass-functions-empirical}) above the free-streaming scale $M_{\rm fs}$ gives a prediction of $1380$ objects in the simulation volume.


\subsection{Matching \peaks\ and haloes}
\label{subsec:matching}
If the excursion set picture is valid, the \peaks\ we identify should be correlated with the actual proto-haloes. Could the mass function mis-match simply be because our low mass \peaks\ are not associated with proto-haloes? To assess this, we match the proto-halo catalogue and the \peaks\ catalogue as follows.

For every proto-halo, we find the \peaks\ contained inside of a sphere of its Lagrangian radius, and associate the \peak\ of highest mass with the halo. This procedure matches $978$ (i.e., $64\%$) of the proto-haloes to \peaks.
We then repeat the same procedure matching \peaks\ to haloes using the filter radius on which the \peak\ was identified, and in this case we can match $935$ (i.e., $74\%$) of the \peaks\ to proto-haloes. (Including the sub-peaks in the analysis makes these numbers $68\%$ and $72\%$, respectively.) We discuss possible reasons for the relatively large fraction of mismatched objects below \citep[see also][]{lp11}. In Appendix~\ref{app:peakshapes}, we also discuss the effect of repeating the exercise with the ellipsoidal collapse barrier \eqref{SMTbarrier}.

As a visual example, we note that the well-behaved proto-halo in the top panel in Figure~\ref{fig:lagrange_examples} was assigned a matching \peak\ while the other two distorted objects were not. While the majority of the proto-haloes we can match to \peaks\ look like the object in the top panel and the majority of unmatched proto-haloes are distorted, there are also a number of examples of well-behaved proto-haloes that are not matched, as well as distorted proto-haloes that are.

The top panel of Figure~\ref{fig:matching_mass_halo2walk} shows the masses of the proto-haloes that we could match to \peaks\ compared to the corresponding \peak\ masses. The points are coloured by the spherically averaged proto-halo sphericity $S$ (equation~\ref{sphericity-def}). There is a strong trend of $S$ with halo mass: low mass haloes are decidedly aspherical. This is consistent with the CDM results of \citet{lp11-collapse}. Additionally, the scatter in mass assignment also correlates strongly with $S$, with low mass, aspherical haloes having the largest scatter. However, at a given halo mass, the mass mismatch \emph{on average} does not seem to correlate strongly with halo shape.

The histograms in the bottom panel of the Figure show the mass function of matched (orange) and unmatched (blue) proto-haloes, with the gray histogram showing the total halo mass function (same as the red histogram in Figure~\ref{fig:mass-functions-empirical}). The matched fraction is quite large at the highest masses (reaching $100\%$ for $m\gtrsim3\times10^{13}\Ms$), remains approximately constant at intermediate masses $m\sim M_{\rm hm}$ and falls significantly at low masses $m\lesssim 10^{\rm 12}\Ms$.

In the top panel of Figure~\ref{fig:matching_mass_walk2halo}, we show the ratio of \peak\ mass to proto-halo mass as a function of \peak\ mass, for \peaks\ that we could match to haloes, coloured in this case by the spherically averaged proto-halo curvature $x$. We see a weak trend of curvature with mass mismatch: \peaks\ matching shallower proto-haloes appear to have larger mass mismatches.

The histograms in the bottom panel of the Figure show the mass function of matched (orange) and unmatched (blue) \peaks, while the gray histogram is the same as in Figure~\ref{fig:matching_mass_halo2walk} and shows the mass function of all haloes. We clearly see that most of the unmatched \peaks\ were assigned dramatically lower masses than any proto-halo in the sample. This could indicate that the ESP picture is, in fact, not appropriate for these objects. However, the presence of a small fraction of low mass \peaks\ that \emph{do} have matching proto-haloes, which in turn have \emph{larger} true masses and low curvatures, suggests that the explanation of this trend could be more subtle. We therefore explore the properties of the mismatched objects in more detail in the next subsection. Recall that the sum of the mass functions of matched and unmatched \peaks\ is consistent with analytical ESP predictions (compare the smooth black curves and solid orange histogram in Figure~\ref{fig:mass-functions-empirical}).


\subsection{Failures of the proto-halo $\leftrightarrow$ \peak\ matching}

\begin{figure}
\begin{center}
\includegraphics[width=0.4\textwidth]{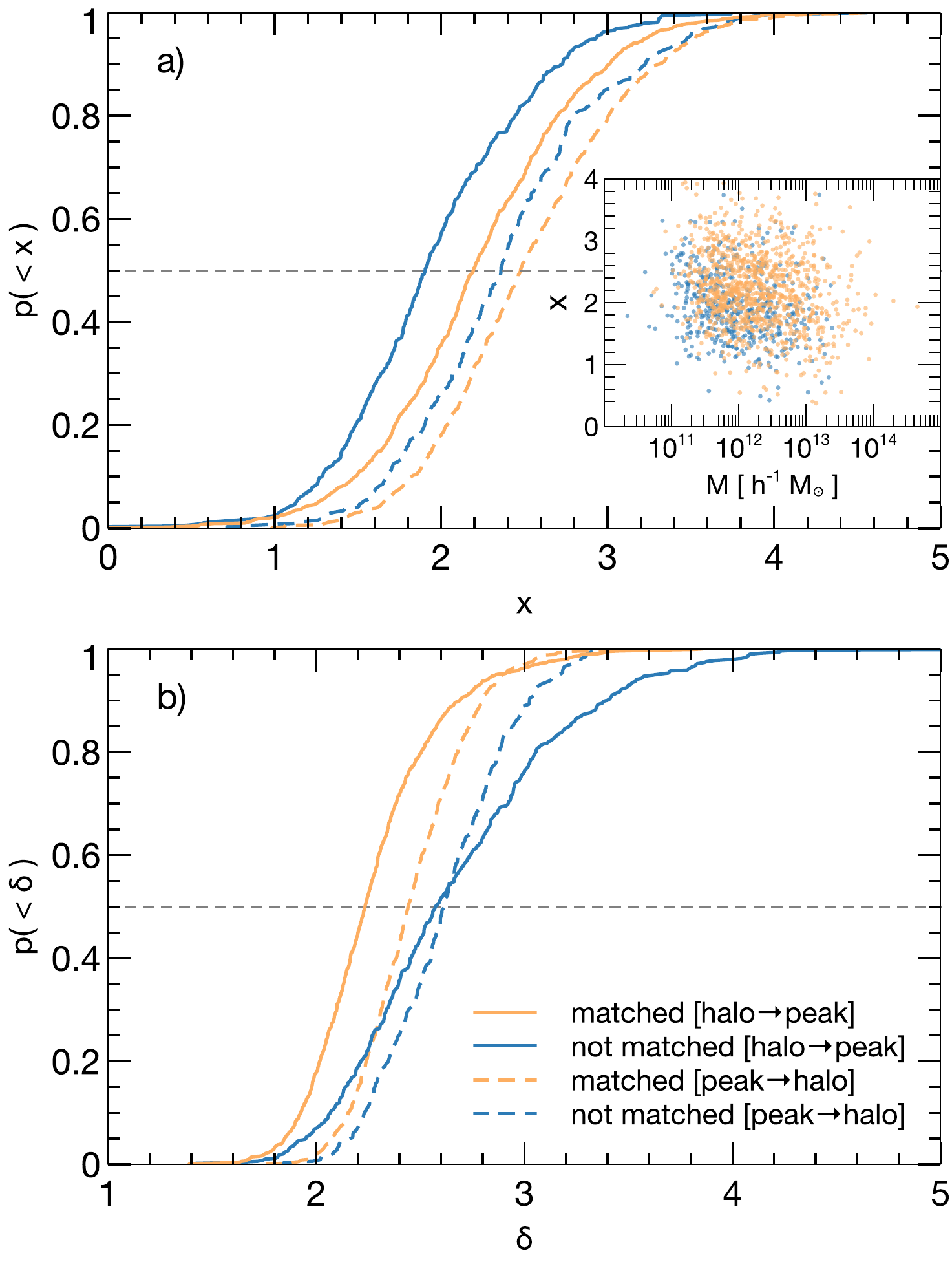}
\end{center}
\caption{Cumulative distributions of spherically averaged peak curvature (top) and density (bottom) for the proto-haloes that could be matched to \peaks\ (solid orange) and for proto-haloes for which no corresponding \peak\ could be found (solid blue). The dashed orange (blue) curves show the corresponding quantities for \peaks\ that could (not) be matched to proto-haloes. The inset in the top panel shows the joint distribution of proto-halo mass and curvature in the matched (orange) and unmatched (blue) cases. The the absence of a significant mass dependence of $x$ shows that the trend seen in the cumulative distribution -- unmatched objects have smaller curvatures -- is not entirely caused by the preferentially low masses of the unmatched objects.
}\label{fig:ecdf_curvature}
\end{figure}

We have already seen above that the masses of the \peaks\ that cannot be matched to proto-haloes are too low compared to the mass function of haloes. As illustrated in Figure~\ref{fig:ecdf_curvature}, these unmatched proto-haloes tend to have  significantly lower peak curvatures and higher densities (solid blue curves in the top and bottom panels, respectively) than those that could be matched to \peaks\ (solid orange curves in the respective panels). Similar trends are also seen, albeit to a lesser extent, in the densities and curvatures of \peaks\ that (do not) have matching proto-haloes, as seen in the dashed orange (blue) curves in the two panels. The inset in the top panel of the Figure shows the joint distribution of curvature and mass for the matched (orange) and mismatched (blue) proto-haloes. The absence of a significant mass dependence of $x$ indicates that the difference in peak curvature is not a result of the two populations occupying different mass ranges. 

\begin{figure}
\begin{center}
\includegraphics[width=0.4\textwidth]{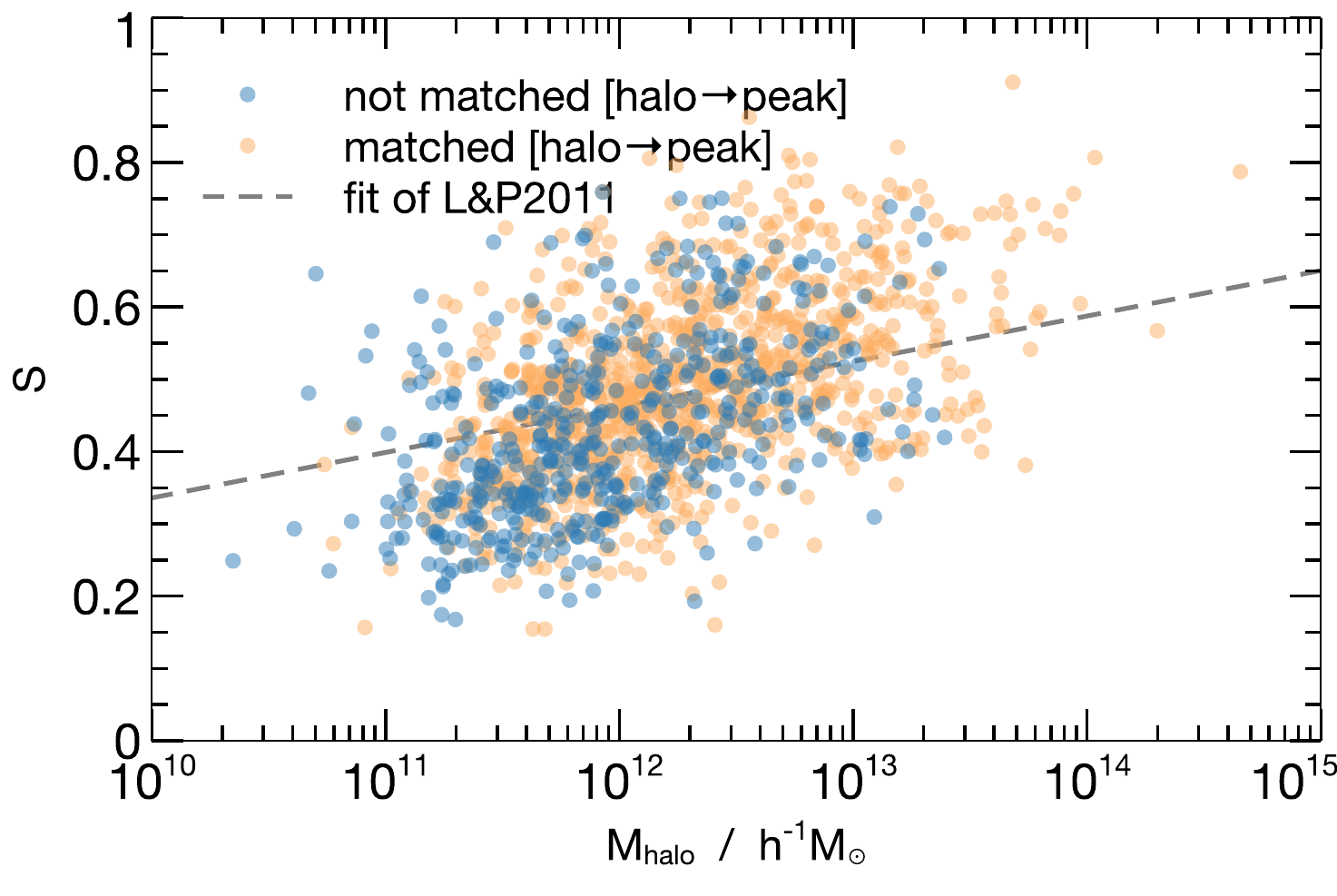}
\end{center}
\caption{Joint distribution of proto-halo mass and sphericity $S$ (equation~\ref{sphericity-def}) in the matched (orange) and unmatched (blue) cases. There is a strong trend of sphericity with mass, as noted in Figure~\ref{fig:matching_mass_halo2walk}. Apart from this, however, there is no significant difference between the sphericities of matched and unmatched proto-haloes. The dashed line, which is somewhat shallower than the measured mass trend, shows the linear fit to corresponding measurements in CDM simulations presented by \citet{lp11-collapse}.}
\label{fig:match_sphericity_mass}
\end{figure}

Figure~\ref{fig:match_sphericity_mass} shows the distribution of mass and sphericity $S$ for the matched (orange) and unmatched (blue) proto-haloes. There is a strong trend of sphericity with mass, as noted in Figure~\ref{fig:matching_mass_halo2walk}. Apart from this, however, there is no significant difference between the sphericities of matched and unmatched proto-haloes. The dashed line, which is somewhat shallower than the measured mass trend, shows the linear fit to corresponding measurements in CDM simulations presented by \citet{lp11-collapse}.


\subsection{Discussion}
\label{subsec:summarynumerics}

Our comparison of the outcomes of simulation and the empirical walks suggests that, at least at masses $m\gtrsim M_{\rm hm}$, the empirical approach more or less correctly predicts both the locations and masses of collapsed haloes. At smaller masses, however, the algorithm is able to predict the location of a collapsed object only for a relatively small fraction ($\sim 55\%$ of proto-haloes below $10^{12}\Ms$ have an associated \peak, and the fraction falls below $50\%$ quickly below this mass scale) and almost always predicts too small a mass in these cases. The large fraction of \peaks\ that cannot be matched to proto-haloes are also predicted to have dramatically lower masses than any proto-halo in the simulation.

We have also seen that, on average, the mass and position mis-matches seem to be uncorrelated with the shapes of the proto-haloes. That is to say, although smaller proto-haloes are decidedly aspherical (with a scatter in mass mismatch that correlates with sphericity), there seems to be no trend between the average sphericity and the ratio $m_{\rm\peak}/m_{\rm halo}$ in the matched cases (Figure~\ref{fig:matching_mass_halo2walk}) and no significant difference between sphericities of matched and unmatched proto-haloes with $m_{\rm halo}<M_{\rm hm}$ (Figure~\ref{fig:match_sphericity_mass}). 

The unmatched proto-haloes do have smaller curvatures than the matched ones. In principle, this could simply be because of their lower masses. However, the absence of a significant mass dependence of $x$ in the inset in the top panel of Figure~\ref{fig:ecdf_curvature} indicates that this is not the case. Additionally, in the case of \peaks\ matched to proto-haloes, the proto-halo curvature correlates with the mass mismatch (Figure~\ref{fig:matching_mass_walk2halo}). 

Note that the objects identified in the simulation of AHA13 were split into different types (see Section~\ref{sec:halofinding}), the most important being ``type-1'' (virialized haloes) and ``type-2'' (objects in late stages of formation). The results above are for the $1522$ ``type-1'' objects, while we have ignored the $438$ ``type-2'' objects. If the latter also form from peaks, our choice of ``type-1'' could be a cause for concern since a peaks-based analysis such as ESP might simply not be able to distinguish between them. Indeed, we do not find a significant difference between the proto-halo regions of ``type-2'' and ``type-1'' objects. Moreover, we find that including the ``type-2'' objects in the analysis on the same footing as ``type-1'' leads to a larger number ($1097$) of matched \peaks, meaning that $87\%$ of our \peaks\ can be matched to \emph{some} object that is either about to or has completely virialized. (The fraction of \emph{unmatched} proto-halo patches is now $\sim44\%$, as compared with $33\%$ when using only ``type-1''. This is largely simply because the combined number of ``type-1'' and ``type-2'' objects ($1960$) is significantly larger than that of the \peaks, which can only be accommodated in the ESP calculation by lowering the collapse threshold.)

Interestingly, AHA13 found that the transition between ``type-2'' and ``type-1'' occurs fast and is associated with a rapid mass growth, bringing a ``type-2'' object to a mass around or above the half-mode mass by the time it has virialized and thus turned into a ``type-1'' object. This is consistent with the picture that power spectra with steeper (effective) slopes show enhanced accretion rates \citep{lc93,lc94}. These observations suggest that the excursion set calculation could be failing because it is unable to capture the quick mass growth that ``type-1'' objects experience around the half-mode mass scale, possibly due to an incorrect prediction of \emph{collapse-time} for a given peak-patch. The rapid transition between these two types of objects means that even small errors in predicting the collapse-time could dramatically alter the predicted locations and masses of fully virialized haloes.
As we discuss in the next Section, such an error can also account for the correlations we find between proto-halo curvature and mass/location mismatches\footnote{The unmatched proto-haloes also have significantly larger densities than the matched ones. As an additional direct test of the barrier hypothesis, we have performed walks centred at the \emph{known} proto-halo centres. This gives us another catalog of masses and corresponding Lagrangian properties, and removes some of the ambiguity associated with off-centring effects which are one potential cause of the low matched fraction we reported above. When using this algorithm, we find that almost all the proto-haloes that were unmatched as per our earlier algorithm are now assigned masses significantly \emph{larger} than their true mass. This is consistent with their larger overdensities compared to the matched proto-haloes: larger local overdensities imply that walks centred at these locations will cross the excursion set barrier at larger mass scales. There is, however, no obvious reason for this trend, and we return to this point in Section~\ref{sec:discussion}.}.


\section{Analytical Results}
\label{sec:analytical}
In this section we use the results of our numerical study to motivate an analytical approximation which captures the sharp cut-off in the mass function better than the standard ESP calculation.

\subsection{A possible explanation for the mis-match between \peaks\ and proto-haloes}
\label{subsec:possibleexplanation}
The behaviour discussed above might be explained if, at small masses, the algorithm systematically overpredicts the \emph{time} at which a given peak-patch should collapse. This is because a patch that collapses earlier than predicted will have time to accrete mass by the time of interest, and will consequently have a larger mass than predicted. As noted by AHA13, low mass WDM haloes tend to grow much more rapidly than their high mass counterparts, so even a small error in collapse-time could have a dramatic impact on the predicted mass function. 
This is further corroborated by our observation in Section~\ref{subsec:matching} that the assignment of peaks to either virialized haloes or objects in the late stages of formation is somewhat uncertain.

Let us suppose that there is in fact such a systematic uncertainty in collapse-time. This is not an unreasonable assumption; similar effects have been noticed and discussed by other authors \citep{Monaco99,gmst07} in the case of CDM. Such effects could arise due to simplifying choices made in models such as ellipsoidal collapse (see Appendix~\ref{app:ecd} for a justification),  as well as due to other physical mechanisms such as assembly-bias\footnote{It is also worth noting that \citet{Monaco99} discussed the difference between what he called orbit-crossing (first-axis collapse) and multi-streaming (last-axis collapse). Standard ellipsoidal collapse models employ the latter as the criterion for collapse, and Monaco argued why one might then expect to correctly predict the locations of collapse but not the halo masses. In particular, he argued that orbit crossing may be a better indicator of halo mass. The semi-analytic code {\sc Pinocchio} \citep{pinocchio,pinocchio-reloaded} uses orbit-crossing as a key ingredient in halo identification, and as a follow-up it would be very interesting to check how accurately {\sc Pinocchio} describes the mass function cut-off in WDM cosmologies.}.
Can this explain the orders-of-magnitude mass increases that are required to go from the \peaks\ mass function to the halo mass function? To see why this is indeed the case, consider the following.

\begin{figure}
\begin{center}
\includegraphics[width=0.35\textwidth]{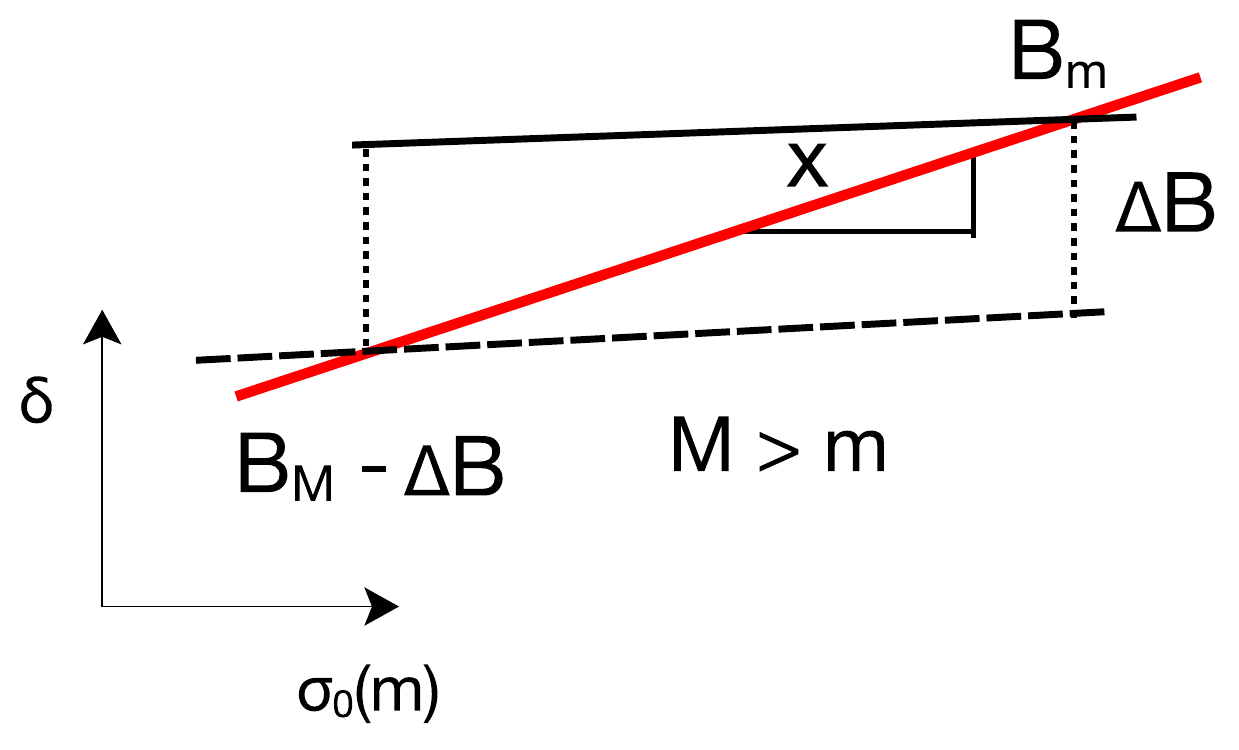}
\end{center}
\caption{An illustration of mass reassignment: An \peak\ is identified at scale $m$ with height $\del=B_m$ and curvature $x$. If the collapse-time prediction in the dynamical model is a systematic overestimate, the \peak\ is allowed to accrete mass for a time $\Del t$ which translates into a lowering of the barrier by $\Del B$. Since the random walk describing the \peak\ has strongly correlated steps, the accretion occurs along essentially a straight line of slope $\dot\del=x/\gam$ in the $\del$-$\sig_0$ plane. This sets the mass scale $M>m$ which is ultimately assigned to this \peak.}
\label{fig:reassign_mass}
\end{figure}

A collapse-time uncertainty can be interpreted as introducing a \emph{second} barrier in the problem, in the sense that if a patch has been identified at mass $m$ using the barrier $B$ (in this case the one in equation~\ref{simplebarrier}) it must then be allowed to accrete until mass $M$ when its density $\del_{\rm patch}$ reaches the ``mass re-assignment'' barrier $B_{\rm mra} = B-\Del B$, where $\Del B$ is positive if the original collapse-time prediction was an overestimate. Figure~\ref{fig:reassign_mass} illustrates the point. Since TopHat/Gaussian filtering induces strong correlations between walk heights, the scale $M$ at which $\del_{\rm patch}$ reaches $B_{\rm mra}$ is essentially determined by the walk height and slope at scale $m$. The height at $m$ is just $B(m)$, and in the excursion set peaks picture the walk slope is strongly correlated with the peak curvature, so that $\der\del/\der\sig_0 \approx x/\gam$ (this relation is exact for Gaussian filtering). The model therefore says that sharper peaks will tend to have more closely matched masses, while shallower peaks will have larger mismatches. 

Note also that the WDM transfer function leads to a ``freeze-out'' of all power spectrum integrals as $m\to0$ (c.f. Section~\ref{subsec:theoryexpect}). This will amplify the above effect at small masses where a small change in $\sig_0$ will imply a huge change in $m$. Additionally, peaks of lower significance will tend to be shallower on average, and this will also systematically enhance the mismatch at low masses.

If this idea is correct, then we should see two effects. Firstly, matched proto-halo patches with lower curvatures should have preferentially larger mass mismatches and vice-versa; and secondly, unmatched proto-haloes must have preferentially low values of curvature (a shallow walk that ``freezes'' before crossing an incorrect barrier will not register as a potential halo). Figure~\ref{fig:matching_mass_walk2halo} is consistent with the first effect, while the second effect is seen quite clearly in Figure~\ref{fig:ecdf_curvature} (see also the discussion in Section~\ref{subsec:summarynumerics}).

In the following, we will therefore assume that the idea of a collapse-time overprediction is correct, and leave for future work a more detailed modelling of the scatter of the mass mismatch by including, e.g., the effect of proto-halo sphericity. We can implement the notion of a second barrier in the ESP calculation as follows. We start by recapitulating the calculation of \citet{psd13}, which we refer to as standard ESP.

\subsection{Standard excursion set peaks}
\label{subsec:stdESP}
The predicted number density of \peaks\ in this calculation can be formally written as
\be
\der n/\der\ln m = \int\Cal{D}{\Xv}\,\Cal{N}_{\rm pk}(\Xv)\, \dir(\ln m - \ln\bar m(\sig_0))\,,
\label{dnESP-formal}
\ee
where the integral is over all relevant variables (e.g., peak density, curvature, shear, etc.) and the function $\Cal{N}_{\rm pk}(\Xv)$ incorporates the intrinsic (Gaussian) probability of these variables, as well as the peaks constraint \eqref{pk-constraint} and the excursion set constraint which requires first crossing of the chosen barrier. The latter means that the integration variables also include $\sig_0$, and the Dirac delta $\dir(\ln m - \ln\bar m(\sig_0))$ then assigns the mass according to the first-crossing scale $\sig_0$, where $\bar m(\sig_0)$ is the inverse function of $\sig_0(m)$ (see Appendix~\ref{app:espStoch} for details).

In the calculation of \citet{psd13}, the barrier was assumed to be $B=\delc+\beta\sig_0$ where $\beta$ is a stochastic variable whose distribution was motivated by the CDM measurements of \citet{Robertson2009} and was assumed independent of the tensors $\p_{ij}\psi$ and $\p_{ij}\del$. In particular, $p(\beta)$ was taken to be Lognormal with mean $0.5$ and variance $0.25$. The resulting mass function is then \citep[equations~12 and~13 of][]{psd13}
\begin{align}
\der n_{\rm ESPstd}/\der\ln m &=  \nu\,\Cal{N}_{\rm ESPstd}(\nu)\,\left|\der\ln\sig_0/\der\ln m\right|\,,
\label{dnESP-std}
\end{align}
with $\nu\equiv\delc(z)/\sig_0(m)$ and
\begin{align}
\nu\,\Cal{N}_{\rm ESPstd}(\nu) &= \int\der\beta\, p(\beta)\frac{{\rm e}^{-(\nu+\beta)^2/2}}{\sqrt{2\pi}\,V_{\ast}}\int_{\beta\gam}^\infty\der x\left(x/\gam-\beta\right)F(x)\notag\\
&\ph{\int(x/\gam-\beta)}\times
p_{\rm G}(x-\beta\gam-\gam\nu;1-\gam^2)\,,
\label{NESP-std}
\end{align}
where $F(x)$ is the BBKS curvature function (equation~\ref{bbks-Fx}) and $p_{\rm G}(y-\mu;\Sigma^2)$ is a Gaussian in the variable $y$ with mean $\mu$ and variance $\Sigma^2$. The solid black curve in \fig{fig:mass-functions-empirical} shows this expression\footnote{Since \citet{psd13} were interested in a CDM mass function, they used a TopHat filter to compute $\sig_0$ but a Gaussian filter for $\sig_1$ and $\sig_2$. (Recall $\sig_2$ diverges for a TopHat filtered CDM spectrum.) The smoothing scale $R_{\rm G}$ for the latter was set by demanding $\avg{\del_{\rm G}|\del_{\rm TH}}=\del_{\rm TH}$, i.e. $\avg{\del_{\rm G}\del_{\rm TH}}=\sig_0(R_{\rm TH})^2$. Consequently, $V_\ast$ was computed using the Gaussian filter and $\gam$ was defined using mixed filtering. We used this prescription for the solid black curve in \fig{fig:mass-functions-empirical}. To keep things simple in the present work, however, we will define \emph{all} quantities in the analytical calculation using Gaussian filtering, with the filtering scale matched to the mass using the relation mentioned above. We have checked that switching to TopHat filtering for defining $\sig_0$ has little impact on our results. Additionally, using Gaussian filtering throughout guarantees self-consistency; e.g., the relation $\der\del/\der\sig_0=x/\gam$, which we use below, is exact in this case.\label{footnote-Gaussfilt}}, using the WDM transfer function \eqref{Tk-WDM}.

In order to implement the barrier \eqref{simplebarrier}, we must account for the correlation between the eigenvalues of the velocity shear $\p_{ij}\psi$ and those of the density Hessian $\p_{ij}\del$. Although this is straightforward in principle, in practice the misalignment between these tensors turns out to be cumbersome to deal with (see Appendix~\ref{app:espStoch}). We therefore explore a simpler, albeit approximate, solution. Following \citet{smt01}, we look for the value at which the distribution $p(Y|\del)$ has its maximum, ignoring the peaks constraint. \citep[This distribution can be obtained by integrating equation A3 of][over the prolateness, and is different from $p(Y,Z=0|\del)$ which is what those authors worked with.]{smt01} This happens at $Y_{\rm max}=0.502\sig_0\approx0.5\sig_0$. We therefore look for the first crossing of the \emph{deterministic} barrier
\be
B=\delc+\bar\beta\sig_0 = \delc+0.5\sig_0\,.
\label{simpledetbar}
\ee
The dashed black curve in \fig{fig:mass-functions-empirical} shows this expression, which amounts to replacing the integral over $\int\der\beta\,p(\beta)$ in \eqn{NESP-std} with the single value $\beta=\bar\beta=0.5$:
\begin{align}
\der n_{\rm ESPdet}/\der\ln m &=  \nu\,\Cal{N}_{\rm ESPdet}(\nu)\,\left|\der\ln\sig_0/\der\ln m\right|\,,
\label{dnESP-det}
\end{align}
with
\begin{align}
\nu\,\Cal{N}_{\rm ESPdet}(\nu) &= \int_{\bar\beta\gam}^\infty\der x\,\Cal{N}_{\rm ESPdet}(\nu,x)\notag\\
&= \frac{{\rm e}^{-\frac12(\nu+\bar\beta)^2}}{\sqrt{2\pi}\,V_{\ast}}\int_{\bar\beta\gam}^\infty\der x\left(x/\gam-\bar\beta\right)F(x)\notag\\
&\ph{\der x(x/\gam-\beta)}\times
p_{\rm G}(x-\bar\beta\gam-\gam\nu;1-\gam^2)\,,
\label{NESP-det}
\end{align}
We see that this describes the dashed blue histogram quite well (this was the result of our empirical walks algorithm for the barrier~\eqref{simpledetbar}, see Section~\ref{subsec:empirical-massfunc}). Consequently, it is also not very different from the solid orange histogram, which was the result of the empirical walks using the stochastic barrier \eqref{simplebarrier}, as well as the standard ESP calculation (solid black curve).

A similar calculation gives the predicted distribution $p(x|{\rm ESP})$ of \peak\ curvature. For the deterministic barrier \eqref{simpledetbar} this is
\be
p(x|{\rm ESPdet}) = \frac{\int\der\ln\sig_0\,\nu\,\Cal{N}_{\rm ESPdet}(\nu,x)}{\int\der\ln\sig_0\int_{\bar\beta\gam}^\infty\der x\,\nu\,\Cal{N}_{\rm ESPdet}(\nu,x)}\,,
\label{px-esp}
\ee
where $\Cal{N}_{\rm ESPdet}(\nu,x)$ was defined in \eqn{NESP-det}. The dashed black curve in Figure~\ref{fig:peak_prop_distribution} shows the result; this is very similar in shape to the measured proto-halo curvature distribution but has a higher mean value.


\subsection{Re-assigning mass}
\label{subsec:reassign}
To implement the mass re-assignment, we modify \eqn{dnESP-formal} by writing
\be
\der n/\der\ln M = \int\Cal{D}{\Xv}\,\Cal{N}_{\rm pk}(\Xv)\,p(\ln M|\Xv)\,,
\label{dnESP-RA-formal}
\ee
where the probability distribution $p(\ln M|\Xv)$ accounts for the mass re-assignment. If the re-assignment were deterministic, this distribution would be a Dirac delta centred on the appropriate re-assigned mass value. Indeed, this is precisely what \eqn{dnESP-formal} does, except that it gets the mass wrong.
In practice, in addition to changing the mass, we allow for some scatter, which is more realistic and also improves the numerical stability of our calculation.

Suppose the standard calculation identifies an \peak\ and assigns it a mass $m$. If the collapse-time uncertainty discussed earlier leads to a barrier shift $-\Del B$, then the strongly correlated nature of the filtered density contrasts \del\ at different smoothing scales means that the mass scale $M$ at the new barrier satisfies
\be
B_M - \Del B = B_m + \frac{x_m}{\gam_m}\left(\sig_{0M}-\sig_{0m}\right)\,,
\label{RAbarrier}
\ee
where the subscript indicates smoothing scale, and we approximated the walk in density $\del(\sig_0)$ as a straight line with slope $\der\del/\der\sig_0 = x/\gam$. If we assume that $\Del B$ is Gaussian distributed with mean $\overline{\Del B}$ and variance $\sig_{\Del B}^2$, and that $B$ is given by \eqn{simpledetbar}, then we have 
\begin{align}
&p(\sig_{0M}|\Xv) \notag\\
&= \int\der\Del B\,p_{\rm G}\left(\Del B-\overline{\Del B};\sig_{\Del B}^2\right)\theta_{\rm H}\left(\sig_{0m}-\frac{\Del B}{x_m/\gam_m-\bar\beta}\right)
\notag\\
&\ph{\int\der\Del B p_G}
\times\dir\left(\sig_{0M}-\sig_{0m}+\frac{\Del B}{x_m/\gam_m-\bar\beta}\right)
\notag\\
&= \theta_{\rm H}\left(\sig_{0M}\right)\left(x_m/\gam_m-\bar\beta\right)
\notag\\&\ph{(x/\gam)}\times 
p_{\rm G}\left(\left(x_m/\gam_m-\bar\beta\right)(\sig_{0m}-\sig_{0M})-\overline{\Del B};\sig_{\Del B}^2\right)
\label{psig0givenX}
\end{align}
where the Heaviside function $\theta_{\rm H}(\ldots)$ ensures that $\sig_{0M}>0$.
The cumulative probability $p(>\ln M|\Xv)$ satisfies
\begin{align}
p(>\ln M|\Xv) &= p(<\sig_{0M}|\Xv) \notag\\
&= \frac12\bigg[\erfc{\frac{(x_m/\gam_m-\bar\beta)(\sig_{0m}-\sig_{0M})-\overline{\Del B}}{\sqrt{2}\,\sig_{\Del B}}}\notag\\
&\ph{\frac12\bigg[\bigg]}
-\erfc{\frac{(x_m/\gam_m-\bar\beta)\sig_{0m}-\overline{\Del B}}{\sqrt{2}\,\sig_{\Del B}}}\bigg]
\label{p>lnM}
\end{align}
The mass function with re-assigned masses then becomes
\begin{align}
\frac{\der n}{\der\ln M} &= \int\der\ln\sig_{0m}\frac{{\rm e}^{-B_m^2/(2\sig_{0m}^2)}}{\sqrt{2\pi}\,V_{\ast m}}
\int_{\bar\beta\gam_m}^\infty\der x\left(x/\gam_m-\bar\beta\right)\notag\\
&\ph{\int_{\beta\gam_m}}\times F(x)p_{\rm G}\left(x-\frac{\gam_m B_m}{\sig_{0m}};1-\gam^2\right)
p(\ln M|\Xv)\,.
\label{dnESP-RA}
\end{align}
Were we to account for the full stochasticity of $B$ using \eqn{simplebarrier}, the expression for the mass function in \eqn{dnESP-RA} would have an additional integral over $Y$, and the integrand, including the distribution $p(\ln M|\Xv)$, would be more complicated. However, since the mass function without mass re-assignment and with $\beta=\bar\beta=0.5$ describes the results of the empirical walks with barrier \eqref{simplebarrier} quite well (c.f. discussion below equation~\ref{simpledetbar}), we will continue to ignore this inherent stochasticity due to the ellipticity $Y$.

\subsection{An explicit example}
The distribution of $\Del B$ will in general depend on the scale $m$ at which the \peak\ is originally identified; we know that at large $m$ the mass assignment is essentially correct, with small scatter, while there is a trend towards underpredicting masses at small $m$. Since there is little guidance from theory for the actual values of $\overline{\Del B}$ and $\sig_{\Del B}$, we have left these as free parameters, except for requiring that they become numerically small for large $m$ or small $\sig_{0m}$. The following is intended as a proof of principle, and we leave a more detailed analysis and estimate of $\overline{\Del B}$ to future work.

We compare $\sig_{0m}$ with the scale $\sig_{0,{\rm turn}}$ which we define as the scale at which the Jacobian between $\sig_0$ and $m$ becomes small. In particular, we set
\be
\left|\der\ln\sig_0/\der\ln m\right|_{\rm turn} = 0.1\,.
\label{sig0turn}
\ee
In practice, for $m_{\rm dm}=0.25$keV this occurs at $m\simeq3.2\times10^{12}\Ms$ which is close to the half-mode mass scale $M_{\rm hm}\simeq3\times10^{12}\Ms$. We have chosen this definition since it remains well-behaved in the CDM limit as well, whereas the half-mode mass goes to zero in that case. The choice of $0.1$ as the threshold in \eqn{sig0turn} is, however, arbitrary.

\begin{figure}
\begin{center}
\includegraphics[width=0.45\textwidth]{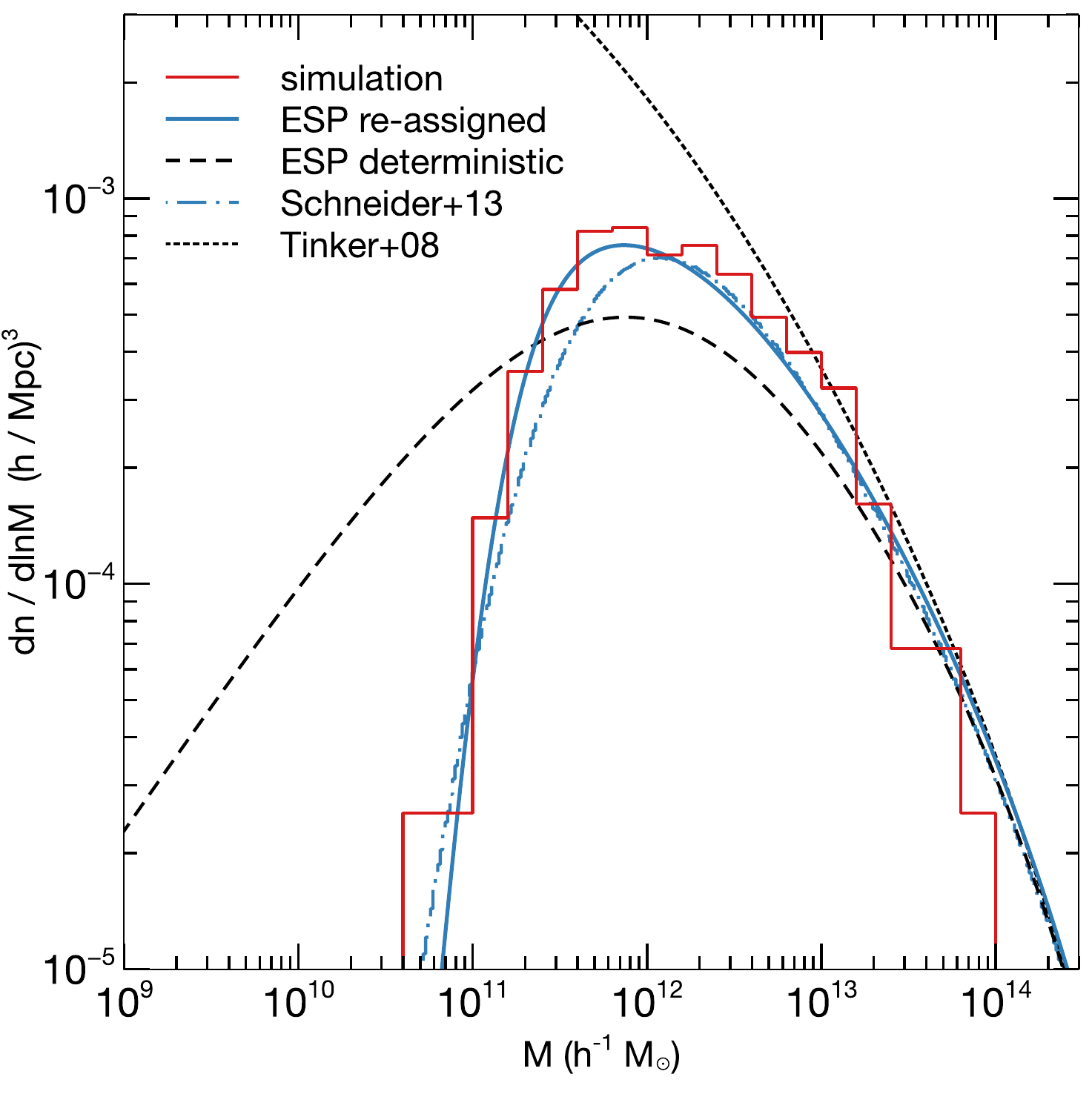}
\end{center}
\caption{Halo mass functions: the solid red histogram and dashed black curve are the same as in \fig{fig:mass-functions-empirical} and show, respectively, the mass function of haloes in the simulation and the ESP prediction \eqref{dnESP-det} using the deterministic barrier \eqref{simpledetbar}. The solid blue curve shows the result of re-assigning masses using \eqn{dnESP-RA}, with parameter values from \eqn{DBchoice}. For comparison, the dot-dashed blue curve shows the fit based on sharp-$k$ filtering presented by \citet[][their spherical collapse fit with $q=1$ and $c=2.7$]{ssr13}, which is a good description of the mass function measured in their simulation. The dotted black curve shows the \citet{Tinker08} fitting form.}
\label{fig:mass-functions-analytical}
\end{figure}

The solid blue curve in \fig{fig:mass-functions-analytical} shows the result of using \eqn{dnESP-RA} after setting
\be
\overline{\Del B} = 5\sig_{\Del B} = 0.175\times(\sig_{0m}/\sig_{0,{\rm turn}})^3\,.
\label{DBchoice}
\ee
The shape of the turnover is quite sensitive to the value of $\overline{\Del B}$, less so to $\sig_{\Del B}$. The numerical values of the amplitude and exponent in the expression on the right are quite degenerate. With these settings, the ESP mass function with re-assigned masses gives a fairly good description of the halo masses in the simulation (histogram). For comparison, the Figure also shows the ESP mass function using the barrier \eqref{simpledetbar} but \emph{before} mass re-assignment (dashed black; this is the same as in \fig{fig:mass-functions-empirical}), and the \citet{Tinker08} fitting function (dotted black). Additionally, the dot-dashed blue curve shows the sharp-$k$ excursion set fit proposed by \citet{ssr13} to their simulations. (For the latter we used their spherical collapse fit, setting their $q=1$ and $c=2.7$, which gives an excellent description of their $z=0$ haloes at $m<10^{15}\Ms$.)

\begin{figure}
\begin{center}
\includegraphics[width=0.4\textwidth]{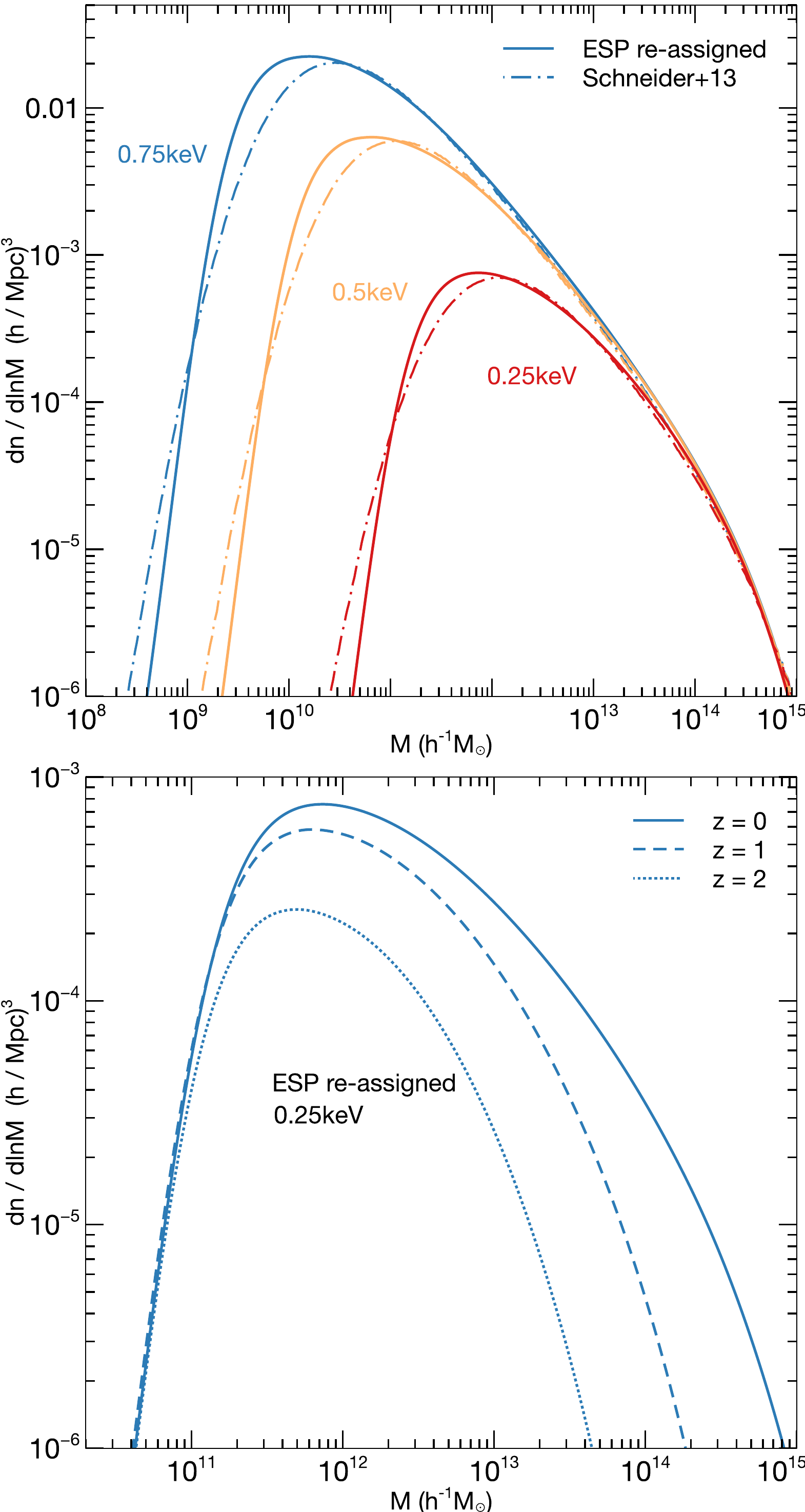}
\end{center}
\caption{(\emph{Top panel}:) Analytical halo mass functions for different values of $m_{\rm dm}$; $0.25$keV (red), $0.5$keV (yellow) and $0.75$keV (blue). The solid curves shows the result of re-assigning masses in the ESP calculation using \eqn{dnESP-RA}, with parameter values from \eqn{DBchoice}. For comparison, the dot-dashed curves shows the fit based on sharp-$k$ filtering presented by \citet[][their spherical collapse fit with $q=1$ and $c=2.7$]{ssr13}, which is a good description of the mass function measured in their simulation. (\emph{Bottom panel}:) ESP mass functions with re-assigned masses at three different redshifts.}
\label{fig:mass-functions-mdmz}
\end{figure}

The top panel of \fig{fig:mass-functions-mdmz} shows our analytical mass function at $z=0$ with masses re-assigned using the same parameter values as in \eqn{DBchoice} for three different values of dark matter particle mass $m_{\rm dm}$ (solid curves). We also show the corresponding curves fit by \citet{ssr13}, again using their spherical collapse fit (dot-dashed curves). In the bottom panel we show our analytical prediction for $m_{\rm dm}=0.25$keV, now for three different redshifts. 
The solid curves marked $0.25$keV in both panels are identical, and the same as the solid blue curve in \fig{fig:mass-functions-analytical}.

We have chosen to model the mass re-assignment on an object-by-object basis, since this is what our empirical walks seem to require. This means that \eqn{dnESP-RA-formal} preserves the total number density of haloes, which is returned as $2.69\times10^{-3}(h/{\rm Mpc})^3$. (As noted earlier, this is somewhat lower than the measured number density of haloes, $2.97\times10^{-3}(h/{\rm Mpc})^3$.) In contrast, the \emph{mass fraction} in collapsed objects given by
\be
f_{\rm coll} = \int\der\ln M\,(M/\bar\rho)\,\der n/\der\ln M\,,
\label{fcoll}
\ee
is not held fixed during the re-assignment, which is obvious since the calculation allows haloes to accrete more mass than is predicted in the standard ESP treatment. 
The values for $f_{\rm coll}$ returned by the analytical calculation before and after mass re-assignment are, respectively, $0.19$ and $0.23$. In comparison, the mass fraction in actual haloes is $0.18$, but note that this number can fluctuate due to sample variance effects at the high-mass end.


\subsection{Consequences for CDM}
\label{subsec:cdm}

\begin{figure}
\begin{center}
\includegraphics[width=0.4\textwidth]{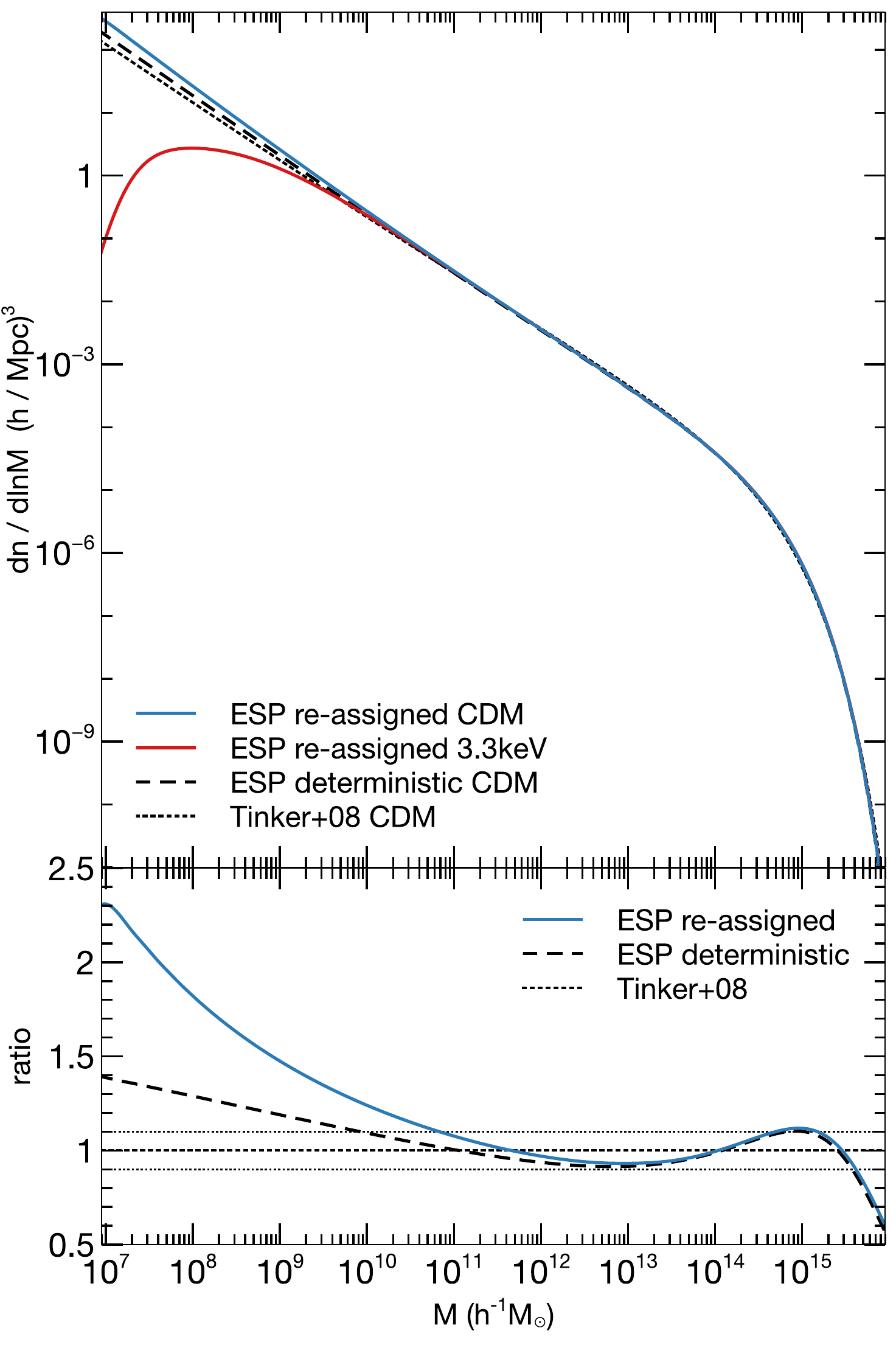}
\end{center}
\caption{ESP mass function predictions for CDM using the deterministic barrier \eqref{simpledetbar} (dashed black) and after re-assigning masses (solid blue). The dotted black curve shows the \citet{Tinker08} fitting form, which was calibrated over a range between $\sim10^{10.5}\Ms$ and $\sim10^{15.5}\Ms$. The solid red curve shows the prediction, after mass re-assignment, for a ``realistic'' WDM particle candidate with $m_{\rm dm}=3.3$keV. In all cases the mass re-assignment parameters are the same as in \eqn{DBchoice}. The lower panel shows the ratio of the CDM predictions to the Tinker et al. fit. 
}
\label{fig:mass-functions-cdm}
\end{figure}

The predictions of the ellipsoidal collapse model, augmented by a systematic uncertainty in mass assignment, accurately describe the WDM mass function. By the logic discussed earlier, the same expressions with the CDM transfer function should describe the CDM mass function. 
In particular, the half-mode mass scale for CDM is small enough that, in practice, every halo has a virialized progenitor. This means that the effects of collapse-time uncertainty -- which were very pronounced around the half-mode mass of WDM due to the rapid growth of those objects -- are now essentially an uncertainty in the time of major mergers, and consequently the associated mass mismatch must be significantly smaller.
We see in \fig{fig:mass-functions-cdm} that this is indeed the case for masses $m\gtrsim10^{11}\Ms$. 
It is also reassuring to note that changing the value of $\overline{\Del B}$ in the CDM case has much less effect on the mass function at $m\gtrsim10^{11}\Ms$ than in the WDM case. E.g., we have checked that increasing the amplitude in \eqn{DBchoice} by a factor $1.5$ or changing the exponent from $3$ to $2$ both lead to $\lesssim3\%$ changes in the CDM mass function for $m\gtrsim10^{11}\Ms$. 

At lower masses our specific implementation of mass re-assignment predicts a factor $\sim2$ larger number of haloes than expected from the mass function fit by \citet{Tinker08}. This behaviour of the re-assigned mass function at low masses is not very robust, however; it is sensitive to the specific numerical choice in \eqn{sig0turn}. It is possible to adjust this number (and those in equation~\ref{DBchoice}) to simultaneously get a good match to the CDM and WDM simulations, although we have not pursued this exercise here. One must also keep in mind that the \citet{Tinker08} fit was calibrated for masses between $\sim10^{10.5}\Ms$ and $\sim10^{15.5}\Ms$.

In other words, our proposed modifications to the ESP calculation not only correctly describe the sharp turn in the WDM mass function, but also describe the CDM mass function with the same accuracy as the standard ESP calculation of \citet{psd13}. For CDM we have also checked that the linear Lagrangian halo bias predicted by this model (not shown) matches measurements in CDM simulations \citep{Tinker10} with the same accuracy as the \citet{psd13} calculation. However, as noted earlier, low mass in WDM does not mean low significance, and, in principle, low significance CDM haloes could be different from low mass WDM haloes. Testing this would need high resolution CDM simulations, or WDM simulations with a slightly larger mass such as $m_{\rm dm}\sim0.5$keV. 


\section{Discussion and Conclusions}
\label{sec:discussion}

Can we predict where and when haloes form? In this paper, we have thoroughly evaluated our ability to predict the abundance of collapsed objects by performing an in-depth analysis of the properties of the initial density field at the locations where collapse occurs in numerical simulations. To accomplish this, we used a perturbation spectrum with a small-scale cut-off such as those arising in WDM cosmologies. As discussed in Section~\ref{sec:massfunc}, the resulting suppression of low mass haloes provides powerful additional leverage which is absent in the CDM case.

Numerical simulations have traditionally had great difficulty in making a prediction for the abundance of haloes in such a scenario due to the artificial fragmentation of filaments -- a problem that has only recently been overcome by \citet[][AHA13]{Angulo2013}. As a consequence, we were in the unique situation of being able to perform a thorough comparison between this numerical experiment and the mass function predicted from excursion set theory, by analysing the properties of haloes on an object-by-object basis. We summarize our results below, and discuss some outstanding issues.

It is well known that the standard excursion set approach predicts a mass function that is completely inconsistent with the numerical results \citep[see, e.g.,][and also our Figure~\ref{fig:mass-functions-empirical}]{Schneider2012}. We showed that the inclusion of the peaks constraint in excursion sets \citep{ms12,ps12,psd13} leads to a turn-over in the mass function as well as an overall number of collapsed objects that is consistent with the simulation results. However, it also predicts masses around and below the half-mode mass scale that are significantly smaller than those measured in the simulation, leading to a small-mass slope of the mass function ($\der n/\der\ln m \sim m^{2/3}$) that is inconsistent with that found from the simulation (c.f. Figure~\ref{fig:mass-functions-empirical}). This prediction is remarkably robust against changing details of the calculation such as the shape and stochasticity of the barrier.

We next investigated the origin of this discrepancy between simulation and theoretical predictions. In particular, we analysed the Lagrangian properties of ``proto-haloes'' (the initial locations of groups of particles which are eventually identified as haloes in the simulation), and also performed empirical excursion set peak walks in the initial density field used in the simulation. We can summarize our findings as follows:
\begin{enumerate}[1.]
\item All haloes in the simulation are consistent with forming near peaks in the initial density field (Figure~\ref{fig:peak_prop_distribution}).
\item The overdensities of proto-haloes are strongly correlated with their shear ellipticities, but show no correlation with the shear prolateness (Figure~\ref{fig:peakdensities}). The former is expected from arguments based on ellipsoidal collapse dynamics, while the latter is not (compare equation~\ref{simplebarrier} with equation~\ref{SMTbarrier}).
The fact that the proto-halo overdensity has no correlation with its prolateness is intimately connected with the distribution of individual shear eigen-values and hence with the dynamical ordering of the collapse times of each \citep{lp11-collapse,dts13}. It will be interesting to find a dynamical model that is consistent with our results, perhaps along the lines presented by \citet{lp11-collapse}.
\item The number of ``\peaks'' ($1261$) identified by our empirical algorithm (Section~\ref{sec:empirical_walks_method}) is reasonably close to the actual number of proto-haloes ($1522$). 
\item A significant fraction ($74\%$) of \peaks\ can be matched to actual proto-haloes, while $64\%$ of the proto-haloes can be matched to \peaks\ (details in Section~\ref{subsec:matching}). 
\item The curvatures of these matched objects are significantly higher, and their overdensities significantly lower, than those of proto-haloes and \peaks\ that could \emph{not} be matched to each other (Figure~\ref{fig:ecdf_curvature}).
\item Most strikingly, the masses of \peaks\ are systematically lower than the proto-halo masses. This is true for both matched and unmatched objects (respectively, top and bottom panels of Figure~\ref{fig:matching_mass_walk2halo}). Since the \peak\ mass function is very well described by the ESP calculation (Figure~\ref{fig:mass-functions-empirical}), this fully accounts for the discrepancy between the ESP halo mass function and that measured in the simulation.
\item For matched objects, the mismatch in mass assignment correlates with proto-halo curvature (Figure~\ref{fig:matching_mass_walk2halo}), while the scatter in the mass mismatch correlates with proto-halo shape (Figure~\ref{fig:matching_mass_halo2walk}).
\item We have checked that, apart from having larger overdensities and lower curvatures than their matched counterparts, the \emph{unmatched} proto-haloes do not appear to be special in any other property related to the velocity shear, density Hessian or moment of inertia.
\item If we also include in the analysis objects in late stages of formation (``type-2'' in AHA13), $87\%$ of the \peaks\ can be matched to proto-haloes (although the fraction of unmatched proto-haloes is now larger, largely because the total number of proto-haloes increases). These ``type-2'' objects are known to be undergoing rapid mass growth (AHA13).
\end{enumerate}

\noindent Based on these results, we argued that the likely cause for the observed mass mismatch between \peaks\ and proto-haloes is a systematic overprediction of the collapse-time for a given perturbation. We then showed how such an uncertainty can be accounted for and corrected in the excursion set language (Section~\ref{subsec:possibleexplanation} and Figure~\ref{fig:reassign_mass}), and presented an explicit example of such a correction which describes the numerical WDM results very well (Figures~\ref{fig:mass-functions-analytical} and~\ref{fig:mass-functions-mdmz}). 
As an important consistency check, we also showed that the same model gives an accurate description of the \emph{CDM} mass function, with the simple replacement of the WDM initial power spectrum with that of CDM (Figure~\ref{fig:mass-functions-cdm}).

We emphasize that our solution works because it explicitly alters the mass assignment step of the ESP calculation, in our case by introducing a second barrier. Simply introducing new statistical variables defined by smoothing the initial density field in a single-barrier calculation (say, by setting $B\to\delc+Y+x\sig_0$) would not work, because the predicted mass function in this case would still behave as $\der n/\der\ln m \sim m^{2/3}$ at low masses, as discussed in Section~\ref{subsec:theoryexpect}. At the heart of this issue is the difference between the physics of individual halo formation and the statistics of the initial density field: mismatches in collapse time predictions are primarily a physical, not statistical, problem. In CDM, since the $\sig_0(m)$ relation is always steep, errors in the physical collapse model can be accommodated by altering the statistical modelling (e.g., by changing the barrier shape as a function of $\sig_0$, or by introducing stochasticity in the barrier). WDM, on the other hand, presents us with a situation where such solutions no longer work since the $\sig_0(m)$ relation ``freezes out'' at small masses (\fig{fig:vol}). A full solution of the problem would likely involve a single barrier with an explicit dependence on mass (rather than $\sig_0$) which is fixed by an accurate model of collapse.

Our work can be extended in several directions which could yield clues towards building a more predictive model.
Although we argued that uncertainties in collapse-time can easily arise in toy models such as ellipsoidal collapse, or due to assembly-bias-like effects, we have not provided conclusive evidence pinpointing a specific physical mechanism. 
Furthermore, the proto-halo curvatures we measure are significantly lower than the simplest ESP prediction (Figure~\ref{fig:peak_prop_distribution}). While this might be partially accounted for by the mass re-assignment, it is not clear whether there is a deeper reason for this discrepancy. 
Also, the scatter in \peak\ mass at fixed halo mass for matched objects correlates strongly with the proto-halo shapes (Figure~\ref{fig:matching_mass_halo2walk}). Our mass re-assignment currently incorporates only curvature, and it will be interesting to additionally account for proto-halo shapes. 
The overdensities of unmatched proto-haloes have a significantly higher tail than that of matched objects (Figure~\ref{fig:ecdf_curvature}). This could be indicating that low mass WDM haloes can form near density peaks without necessarily satisfying the excursion set peaks constraints; e.g., there could be some other criterion for a sufficiently overdense patch to virialize. One way forward would be to analyse the local environments of these unmatched objects to look for peculiarities.
Finally, while our chosen particle mass of $m_{\rm dm}=0.25$keV is untenably warm, any realistic massive dark matter candidate would lead to very similar phenomenology around its half-mode mass scale, and our analysis would be relevant whenever this scale can be numerically resolved. As mentioned earlier, however, for a particle much colder than our present choice, low mass haloes will also have low \emph{significance} ($\nu\ll1$) which was not possible in our case, and this could lead to additional effects.
Clearly, it will be of great interest to investigate these aspects further in future work.


\section*{Acknowledgements}
We thank Raul Angulo for providing us with the halo catalogues published in AHA13. 
We wish to thank him, Ravi Sheth, Tom Abel and Cristiano Porciani for insightful discussions and comments on the draft.
O.H. acknowledges support from the Swiss National Science Foundation (SNSF) through the Ambizione fellowship.



\appendix
\section{Fully stochastic excursion set peaks}
\label{app:espStoch}
Here we present some of the formal details of the excursion set peaks calculation. This will highlight the difficulty in working with the full stochasticity that must be dealt with when using a barrier such as \eqref{simplebarrier}, and motivate the simpler, deterministic approximation \eqref{simpledetbar} used in the main text. 

\subsection{Formal expression for mass function}
In what follows, an overdot denotes a derivative with respect to $\sig_0$. All variables are assumed to be Gaussian filtered on a scale $R$ which is related to the mass $m$ through $m=(4\pi/3)\bar\rho R_{\rm TH}^3$ where $R_{\rm TH}$ satisfies $\avg{\del_{\rm G}(R)\del_{\rm TH}(R_{\rm TH})} = \avg{\del_{\rm TH}(R_{\rm TH})^2}$ with the subscripts `G' and `TH' on \del\ denoting Gaussian and TopHat filtering, respectively \citep[][see also footnote~\ref{footnote-Gaussfilt}]{psd13}. In practice this gives $R\approx0.46R_{\rm TH}$ with a slow variation. We will drop the filter subscripts below.

The ESP mass function assuming a barrier $B$ (which can be stochastic) can be written as \eqn{dnESP-formal} where, in full glory, we have
\be
\int\Cal{D}\Xv \equiv \int\der\ln\sig_0\,\der^6\psi\,\der^6\zeta\,\der^3\eta
\label{DX}
\ee
and
\be
\Cal{N}_{\rm pk}(\Xv) \equiv p(\p_{ij}\psi,\p_{ij}\del,\nab\del)\, {\rm\bf Pk}(\p_{ij}\del,\nab\del)\,{\rm\bf ES}(\sig_0,\{\del,B\})\,.
\label{Npk-X}
\ee
The expression \eqref{DX} involves a total of $16$ integration variables: the smoothing scale represented by $\sig_0$, the $6$ independent components of the shear tensor $\p_{ij}\psi$ represented by $\der^6\psi$, similarly the components of the Hessian of the density $\p_{ij}\del$ represented by $\der^6\zeta$, and finally the $3$ components of the density gradient $\nab\del$ represented by $\der^3\eta$. Strictly speaking, we must also include the scale derivative of density $\dot\del$ (which will appear in the excursion set constraint) as a separate variable, but Gaussian filtering ensures that $\dot\del=x/\gam$ where $x$ was defined in \eqn{x-def} and \gam\ in \eqn{gam-Vst}, so this is included in $\p_{ij}\del$.

The raw number density of peaks $\Cal{N}_{\rm pk}(\Xv)$ defined in \eqn{Npk-X} consists of the following quantities: the joint (Gaussian) distribution function $p(\p_{ij}\psi,\p_{ij}\del,\nab\del)$ of the shear, density gradient and density Hessian smoothed on scale $R$; the peaks constraint ${\rm\bf Pk}(\p_{ij}\del,\nab\del)$ which enforces $\nab\del=0$ and $\zeta_i<0$ where $\zeta_i$ are the eigenvalues of $\p_{ij}\del$; and the excursion set constraint ${\rm\bf ES}(\sig_0,\{\del,B\})$ which enforces up-crossing\footnote{\citet{ms12,ms13} discuss why up-crossing is a sufficiently accurate approximation to first-crossing for walks with correlated steps.} of the barrier by the random walk at the scale $\sig_0(R)$. The compact notation in ${\rm\bf ES}(\ldots)$ hides the fact that the up-crossing condition will introduce a dependence on $\dot\del=x/\gam$ and possibly other stochastic quantities through $\dot B$. The Dirac delta in \eqn{dnESP-formal} then sets the mass to be $m=\bar m(\sig_0)$ where $\bar m(\sig_0)$ is the inverse function of $\sig_0(R(m))$ as discussed above, which is straightforward to compute numerically.

In detail, following \citet[][hereafter, BBKS]{bbks86}, we have 
\begin{align}
{\rm\bf Pk}(\p_{ij}\del,\nab\del) &= \dir(\nab\del)\,|\zeta_1\zeta_2\zeta_3|\theta_{\rm H}(-\zeta_3)
\label{Pk-constraint}
\end{align}
where we have assumed the ordering $\zeta_1\leq\zeta_2\leq\zeta_3$, while the excursion set constraint can be written as
\be
{\rm\bf ES}(\sig_0,\{\del,B\}) = (x/\gam-\dot B)\,\theta_{\rm H}(x/\gam-\dot B)\,\dir(\mu-B/\sig_0)\,,
\label{ES-constraint}
\ee
where we defined $\mu\equiv\del/\sig_0$. The Dirac-delta in \eqn{ES-constraint} enforces barrier-crossing, while the terms involving $x/\gam=\dot\del$ ensure that this is an \emph{up}-crossing.

The intrinsic Gaussian distribution of these variables couples the tensors $\p_{ij}\psi$ and $\p_{ij}\del$ through the correlation structure \citep{vdwb96}
\begin{align}
\avg{\p_{ij}\psi\,\p_{kl}\psi} &= \frac{\sig_0^2}{15}\left(\del_{ij}\del_{kl}+\del_{ik}\del_{jl}+\del_{il}\del_{kj}\right)\,,\notag\\
\avg{\p_{ij}\del\,\p_{kl}\del} &= \frac{\sig_2^2}{15}\left(\del_{ij}\del_{kl}+\del_{ik}\del_{jl}+\del_{il}\del_{kj}\right)\,,\notag\\
\avg{-\p_{ij}\del\,\p_{kl}\psi} &= \frac{\sig_1^2}{15}\left(\del_{ij}\del_{kl}+\del_{ik}\del_{jl}+\del_{il}\del_{kj}\right)\,,
\label{tensor-correln}
\end{align}
where the $\del_{ij}$ are Kronecker deltas, whereas the vector $\nab\del$ is uncorrelated with the tensors and satisfies
\begin{align}
\avg{\nabla_i\del\,\nabla_j\del} &= \frac{\sig_1^2}{3}\del_{ij}\notag\\
\avg{\nabla_i\del\,\p_{jk}\psi} &= 0 = \avg{\nabla_i\del\,\p_{jk}\del}\,.
\label{vector-correln}
\end{align}

\subsection{Consequences of misalignment}
The twelve degrees of freedom in the two tensors $\p_{ij}\psi$ and $\p_{ij}\del$ are most conveniently organised as the three eigenvalues of each, the three relative Euler angles between their respective eigenvectors, and three additional Euler angles that fix the orientation of one of them with respect to the chosen basis.

The correlation structure between the tensors $\p_{ij}\psi$ and $\p_{ij}\del$ implies that, in a generic realisation of the field, they will be misaligned. I.e., although they are strongly correlated (with correlation coefficient \gam), their respective eigenvectors will not be parallel \citep[see, e.g.,][]{Desjacques08,lw10,dts13}. The misalignment is captured by the three relative Euler angles, and these must be marginalised over (the other three angles never appear in the distribution due to statistical isotropy, and can be trivially marginalised over.). The effect of this marginalisation is to introduce a nontrivial coupling between the anisotropic combinations of eigenvalues of the two tensors. 
So if we define $y$ and $z$ as in \eqns{epk-y-def} and~\eqref{ppk-z-def} then, as shown in an elegant calculation by \citet{Desjacques08}, the marginalisation over Euler angles will introduce terms involving products between at least one of $\{(y+z),(3y-z)\}$ and at least one of $\{(Y+Z)/\sig_0,(3Y-Z)/\sig_0\}$ where $Y$ and $Z$ were defined in \eqns{ev-Y-def} and~\eqref{pv-Z-def}, respectively. This is quite different from the coupling between the isotropic variables $x$ and $\mu$ through the familiar Gaussian term $\sim{\rm e}^{-(x-\gam\mu)^2/2(1-\gam^2)}$ that appears in the BBKS calculation.

This anisotropic coupling is irrelevant if we were to ignore the excursion set constraint ${\rm\bf ES}(\ldots)$ in \eqn{Npk-X} or if the barrier $B$ in ${\rm\bf ES}(\ldots)$ did not depend on any of the anisotropic variables $\{y,z,Y,Z\}$ \citep[an example would be the barrier used by][]{psd13}, since in these cases one could marginalise over, e.g., $Y$ and $Z$ to recover expressions similar to those analysed by BBKS (note that ignoring the excursion set constraint altogether would give back the BBKS calculation exactly). In the case of the barrier \eqref{simplebarrier}, however, the calculation will involve integration of a term like $\sim p(\lam|\zeta){\rm e}^{-(\mu+Y/\sig_0-\gam x)^2/2(1-\gam^2)}$ over $Y$ and $Z$, where the distribution $p(\lam|\zeta)$ of the eigenvalues of $\p_{ij}\psi$ given the eigenvalues of $\p_{ij}\del$ is given by equation~(22) of \citet{Desjacques08}.

\begin{figure}
\begin{center}
\includegraphics[width=0.4\textwidth]{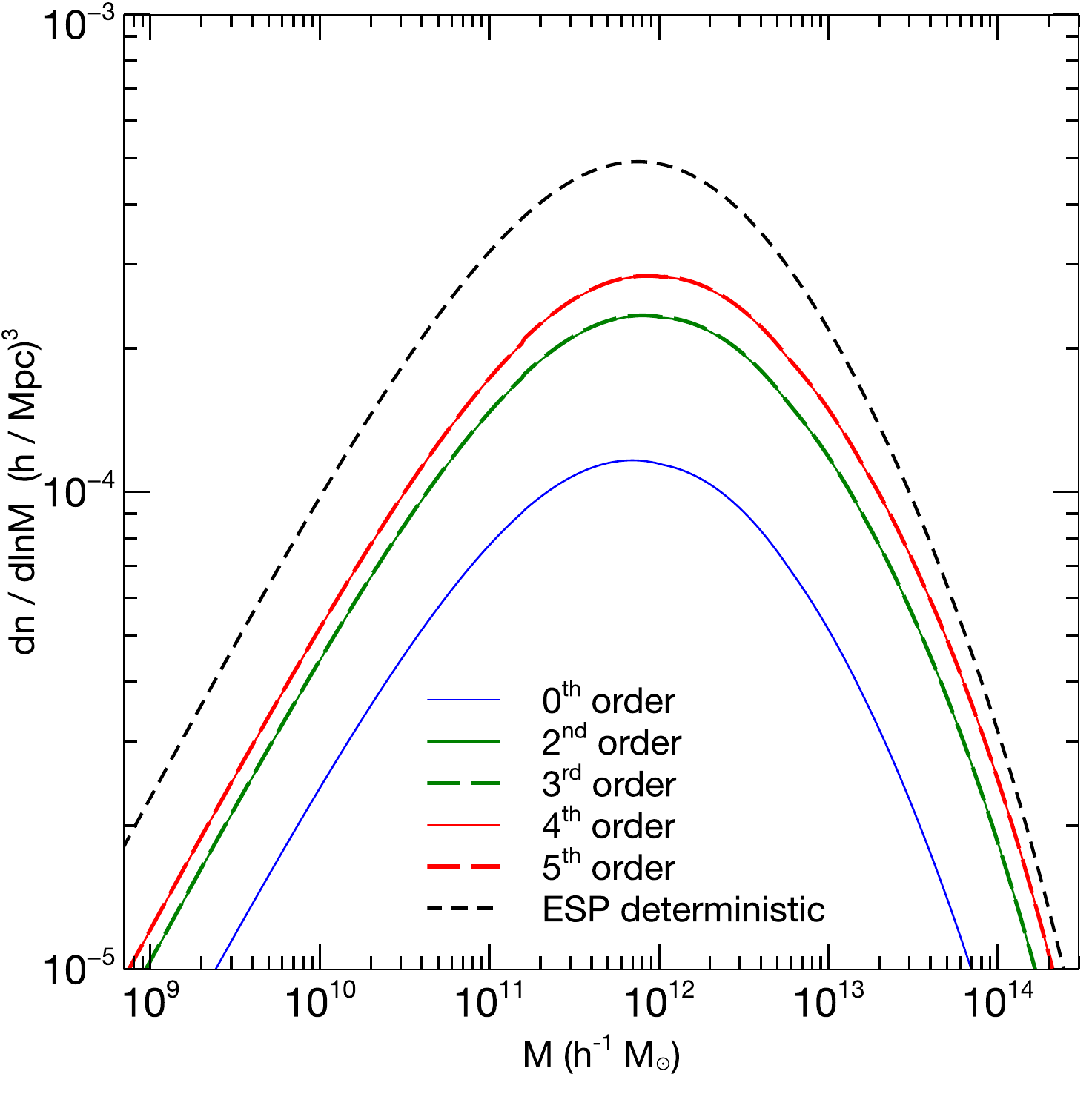}
\end{center}
\caption{Accounting for the effects of stochasticity: ESP mass function predictions for WDM using the stochastic barrier \eqref{simplebarrier} and the approximation $\dot Y = y/\gam$ (see text for details). From bottom to top, the solid curves show the result of keeping the $0^{\rm th}$, $2^{\rm nd}$ and $4^{\rm th}$ order terms in \eqn{W-expand}, while the dashed curves (almost indistinguishable from the upper solid curves) include the $3^{\rm rd}$ and $5^{\rm th}$ order terms. The upper short-dashed black curve is the same as in Figure~\ref{fig:mass-functions-empirical} and shows the ESP calculation using the deterministic barrier \eqref{simpledetbar}, which gives a very good description of the \peaks\ mass function (solid orange histogram in Figure~\ref{fig:mass-functions-empirical}) obtained with the stochastic barrier \eqref{simplebarrier}. While the odd order terms do not contribute significantly to the stochastic calculation, the even order terms have clearly not converged -- the $4^{\rm th}$ order term gives a significant contribution compared to the $2^{\rm nd}$ order calculation.
}
\label{fig:appendix-stochdndlnm}
\end{figure}

This is not all, however. The excursion set constraint also involves the derivative $\dot B$, which in this case would lead to a term $\dot Y$. For Gaussian filtering, it is easy to check that the individual components of $\p_{ij}\psi$ and $\p_{ij}\del$ are related by
\be
\der(\p_{ij}\psi)/\der\sig_0 = -(\p_{ij}\del)/(\gam\sig_2)\,,
\label{tensor-derivative}
\ee
whose trace gives the relation $\dot\del = x/\gam$ quoted earlier. If the tensors were perfectly aligned, this would also imply $\dot Y = y/\gam$. In the general case, however, $\dot Y$ depends on the eigenvalues of $\p_{ij}\del$ and the relative Euler angles between the two tensors in a highly nonlinear way. Using the ellipsoidal collapse barrier \eqref{SMTbarrier} would make things even more complicated.

As an example, consider using the barrier \eqref{simplebarrier} with the approximation $\dot Y = y/\gam$ but otherwise assuming a generic misalignment between the tensors. In this case, after a BBKS-like calculation, the mass function (with or without non-standard mass assignment) becomes
\begin{align}
\frac{\der n}{\der\ln m} &= \int\der\ln\sig_0\,\frac{1}{V_\ast}\frac{N}{\gam}(1-\gam)^4\notag\\
&\ph{\times}
\times\int\der\tilde x \der\tilde y \der\tilde z\, \chi(\tilde x,\tilde y,\tilde z) \int\der\tilde Y \der\tilde Z\, \chi(\tilde\nu+\tilde Y,\tilde Y,\tilde Z) \notag\\
&\ph{\times\times}
\times(\tilde x-\tilde y)\tilde F(\tilde x,\tilde y,\tilde z)\,\tilde Y(\tilde Y^2-\tilde Z^2)\,\Cal{W}(\tilde y,\tilde z,\tilde Y,\tilde Z)\notag\\
&\ph{\times\times\times}
\times{\rm e}^{-\frac12(15\tilde y^2+5\tilde z^2)}{\rm e}^{-\frac12(15\tilde Y^2+5\tilde Z^2)}{\rm e}^{-\frac12\tilde x^2(1-\gam^2)}\notag\\
&\ph{\times\times\times\times}
\times{\rm e}^{-\frac12(\tilde\nu+\tilde Y-\gam\tilde x)^2}\,p(\ln m|\Xv)\,,
\label{dndlnm-vincentstyle}
\end{align}
where $N\equiv5^53^4/(2\pi)^2$, $\{\tilde x,\tilde y,\tilde z\}=\{x,y,z\}/\sqrt{1-\gam^2}$, $\{\tilde \nu,\tilde Y,\tilde Z\}=\{\delc,Y,Z\}/(\sig_0\sqrt{1-\gam^2})$, $\tilde F(\tilde x,\tilde y,\tilde z) = \tilde y(\tilde y^2-\tilde z^2)(\tilde x-2\tilde z)\left((\tilde x+\tilde z)^2-9\tilde y^2\right)$, the function $\chi(s,t,u)$ is unity when $-t\leq u\leq t$, $t\geq0$, $s\geq3t-u$ and is zero otherwise, and the function $\Cal{W}(\tilde y,\tilde z,\tilde Y,\tilde Z)$ which captures the effect of marginalising over the relative Euler angles has the Taylor expansion \citep{Desjacques08}
\begin{align}
\Cal{W}(\tilde y,\tilde z,\tilde Y,\tilde Z) &= 1+\frac{\kappa^2}{10}\Cal{W}_2 + \frac{\kappa^3}{105}\Cal{W}_3 + \frac{\kappa^4}{280}(\Cal{W}_2)^2\notag\\ 
&\ph{1+\frac{\kappa^2}{10}\Cal{W}_2+}
+ \frac{\kappa^5}{2310}\Cal{W}_2\Cal{W}_3 + \Cal{O}(\kappa^6)\,,
\label{W-expand}
\end{align}
where $\kappa\equiv5\gam/4$ and 
\begin{align}
\Cal{W}_2 &= 16\left(3\tilde y^2+\tilde z^2\right)\left(3\tilde Y^2+\tilde Z^2\right)\,,\notag\\
\Cal{W}_3 &= 64\tilde z\tilde Z\left(9\tilde y^2-\tilde z^2\right)\left(9\tilde Y^2-\tilde Z^2\right)\,.
\label{W2-W3-def}
\end{align}
The integrals over $\tilde y$, $\tilde z$ and $\tilde Z$ are tedious but can be expressed in closed form. The remaining integrals over $\tilde x$, $\tilde Y$ and $\sig_0$ must be done numerically. Figure~\ref{fig:appendix-stochdndlnm} shows the result of the zeroth order calculation and that of successively including higher powers of $\kappa$, with standard mass assignment. We see that the odd powers do not contribute significantly. The even powers, however, have not converged since the $\kappa^4$ term gives a significant contribution compared to the second order calculation. Presumably one would have to continue the calculation at least to order $\kappa^6$ (if not $\kappa^8$), to get reasonable convergence. 

\subsection{A simpler way out}
As the results of the previous section show, it rapidly becomes very complicated to deal with the full stochasticity inherent in even a simple barrier prescription like \eqn{simplebarrier}. Luckily, approximating this barrier with the deterministic expression \eqref{simpledetbar} leads to an excellent description of the mass function of \peaks\ found using the stochastic barrier \eqref{simplebarrier} (compare the dashed black curve in \fig{fig:mass-functions-empirical} with the solid orange histogram.) 

The calculation proceeds by setting $B=\delc+0.5\sig_0$ and $\dot B = 0.5$ in \eqn{ES-constraint}, which means that the excursion set constraint is independent of $Y$ and $Z$. These variables, together with the problematic relative Euler angles between the tensors, can then be trivially marginalised over, without having to Taylor expand as in \eqn{W-expand}. The remaining variables can then be dealt with exactly like in BBKS; the only integral that cannot be done analytically is the one over $x$, since this involves the BBKS curvature function $F(x)$ given by
\begin{align}
F(x)&=\frac12\left(x^3-3x\right)\left\{\erf{x\sqrt{\frac52}}+\erf{x\sqrt{\frac58}}
  \right\} \notag\\
&\ph{x^3-3x}
+ \sqrt{\frac2{5\pi}}\bigg[\left(\frac{31x^2}{4}+\frac85\right){\rm
    e}^{-5x^2/8} \notag\\
&\ph{\sqrt{x^3-3x+\frac2{5\pi}}[]}
+ \left(\frac{x^2}{2}-\frac85\right){\rm
    e}^{-5x^2/2}\bigg]\,,
\label{bbks-Fx}
\end{align}
(equations~A14--A19 in BBKS).
This is, of course, exactly the calculation performed by \citet{ps12} and later used by \citet{psd13}. 

The main message here is that the barrier stochasticity induced by ellipsoidal effects is a technical detail that does not address the main problem -- that of mass re-assignment -- we are faced with. 
Rather, the simplicity of the deterministic barrier solution allows us to tackle this problem in a computationally straightforward way, as discussed in the main text. In principle, one could imagine having a physically better motivated model of mass re-assignment; provided it can be expressed in the formal language of \eqn{dnESP-RA-formal}, the hurdles in using it to make mass function predictions would be purely technical.

\section{Ellipsoidal dynamics and collapse-time uncertainty}
\label{app:ecd}
To understand why one might expect a small systematic uncertainty in the predicted collapse time in ellipsoidal dynamics, it will help to first recapitulate some of the basic features of \emph{spherical} collapse. 

Consider the simplest case of an Einstein-deSitter (EdS; $\Omega_{\rm tot}=\Omega_{\rm m}=1$) background. Recall that the comoving Eulerian radius $R$ and overdensity $\Delta = \rho/\bar\rho$ of a collapsing homogeneous sphere with initial overdensity $\del_i=(5/3)\del_0/(1+z_i) > 0$ and comoving Lagrangian radius $R_0$ are given by \citep{gg72}
\begin{align}
\Del = \left(R_0/R\right)^3 &= \frac92\frac{(\theta-\sin\theta)^2}{(1-\cos\theta)^3}\,,\notag\\
\frac{\del_0}{1+z} = \del_0\left(\frac{t}{t_0}\right)^{2/3} = \del_{\rm L} &= \frac35\left(\frac34\right)^{2/3}\left(\theta-\sin\theta\right)^{2/3}\,,
\label{sphcoll-solution}
\end{align}
where $t_0$ is the present epoch and $\del_{\rm L}$ is the linear overdensity extrapolated to redshift $z$. The radius turns around at $\theta=\pi$ and reaches zero at $\theta=2\pi$. In the standard approach, instead of evolving the solution to zero radius, one argues that the radius will become essentially constant once the object virializes. While this is not captured by the spherical model, a simple energy conservation argument says that the physical ``virial radius'' must be one half of the physical radius at turn-around.

The key conceptual point here is that the model itself reaches the virial radius at $t_{\rm vir}=t(\theta=3\pi/2)$ which occurs slightly before the collapse-time $t_{\rm coll} = t(\theta=2\pi)$; in particular, $t_{\rm vir}\approx 0.91\times t_{\rm coll}$. However, one can argue that virialization actually occurs over a dynamical timescale $t_{\rm dyn}$ \citep[the free-fall time at virialization; see, e.g.,][]{paddy93} which, coincidentally, is of order $10\%$ of the collapse time. So in practice it is quite reasonable to use $t_{\rm coll}(\approx t_{\rm vir}+t_{\rm dyn})$ in place of $t_{\rm vir}$ when computing the linearly extrapolated overdensity, which gives the well-known result $\del_{\rm L}=1.686$. \citep[Had we instead evaluated $\del_{\rm L}$ at $\theta=\theta_{\rm vir}$, we would get $\del_{\rm L}=1.583$, a $\sim6\%$ decrement from the traditional value; see, e.g.,][hereafter, BM96]{bm96}. 

The nonlinear overdensity at virialization follows from energy conservation and matching to the turn-around time $t_{\rm ta}$: 
\begin{align}
\Del_{\rm vir} &= \Del_{\rm ta}(t_{\rm vir}/t_{\rm ta})^2(R_{\rm phys,ta}/R_{\rm phys,vir})^3\notag\\
&= (9\pi^2/2)(t_{\rm vir}/t_{\rm ta})^2\notag\\
&\approx (9\pi^2/2)(t_{\rm coll}/t_{\rm ta})^2 = 18\pi^2\,.
\label{Delta-vir}
\end{align}
Notice that the traditional value of $18\pi^2\simeq178$ arises after approximating $t_{\rm vir}\approx t_{\rm coll}$; the actual nonlinear density at $t_{\rm vir}$ would be $\sim20\%$ smaller.

In the ellipsoidal model discussed by BM96 \citep[see also][]{ws79,el95,Monaco99}, it is more difficult to identify conserved quantities since there is no spherical symmetry and nonlinear tides can have a complicated influence. The definition of virialization is therefore somewhat ambiguous. BM96 settled on a simple prescription in which each principle axis is evolved nonlinearly until it shrinks to a predetermined fraction $f_{\rm c}$ of the global scale factor, after which it is assumed to remain fixed at that value. Collapse is defined as the time at which the longest axis (i.e., the smallest shear eigenvalue) satisfies this condition. Choosing $f_{\rm c}=0.178$ ensures that the overdensity at virialization for a spherical configuration is $f_{\rm c}^{-1/3}\simeq178$, the traditional value. Since there is no compensation for dynamical timescales, it seems reasonable that this prescription \emph{underpredicts} collapse times. In the spherical limit, this would be a $\sim 10\%$ effect as discussed above.

The fit \eqref{SMTbarrier} by \citet{smt01} to the resulting linear overdensity values at collapse as a function of ellipticity and prolateness reduces to the traditional spherical collapse result when $Z=0=Y$, meaning that it rescales the BM96 prescription for the barrier by a factor $1.686/1.583=1.06$. This would cause a corresponding $\sim10\%$ increase in collapse time. While this increase is of the correct magnitude in the spherical limit (see above), it is less clear whether this is also true for significantly triaxial configurations. In the EdS case where the growth factor is proportional to the scale factor, one can check that a barrier-rescaling as above does, in fact, rescale the collapse time by a constant at any ellipticity. However, the situation is different, e.g., in flat $\Lambda$CDM with $\Omega_{\rm m} < 1$. In this case, the growth factor is different from the scale factor and one can show that a barrier-rescaling leads to a slightly \emph{larger} effect on the collapse time for $e_{\rm v}>0$ than it does for $e_{\rm v}=0$. In other words, if one were aiming for a $\sim10\%$ increase in collapse time for all ellipticities, then rescaling the barrier by a constant will tend to overcompensate at large ellipticities, and would go in the direction of explaining the results of Section~\ref{subsec:reassign}. 

Of course, the specific prescription for virialization itself is somewhat ad-hoc, and BM96 discuss several modifications, all of which lead to few percent changes in collapse time \citep[see also][]{ab10,lp11-collapse}. Additionally, the specific choice of halo finder in the simulation will also lead to (probably unquantifiable) systematics. It is therefore not unreasonable that uncertainties in dynamical modelling lead to a small (but, in the WDM case, important) systematic overprediction of collapse times.

\section{Peak Shapes and the Collapse Threshold}
\label{app:peakshapes}

\begin{figure}
\begin{center}
\includegraphics[width=0.45\textwidth]{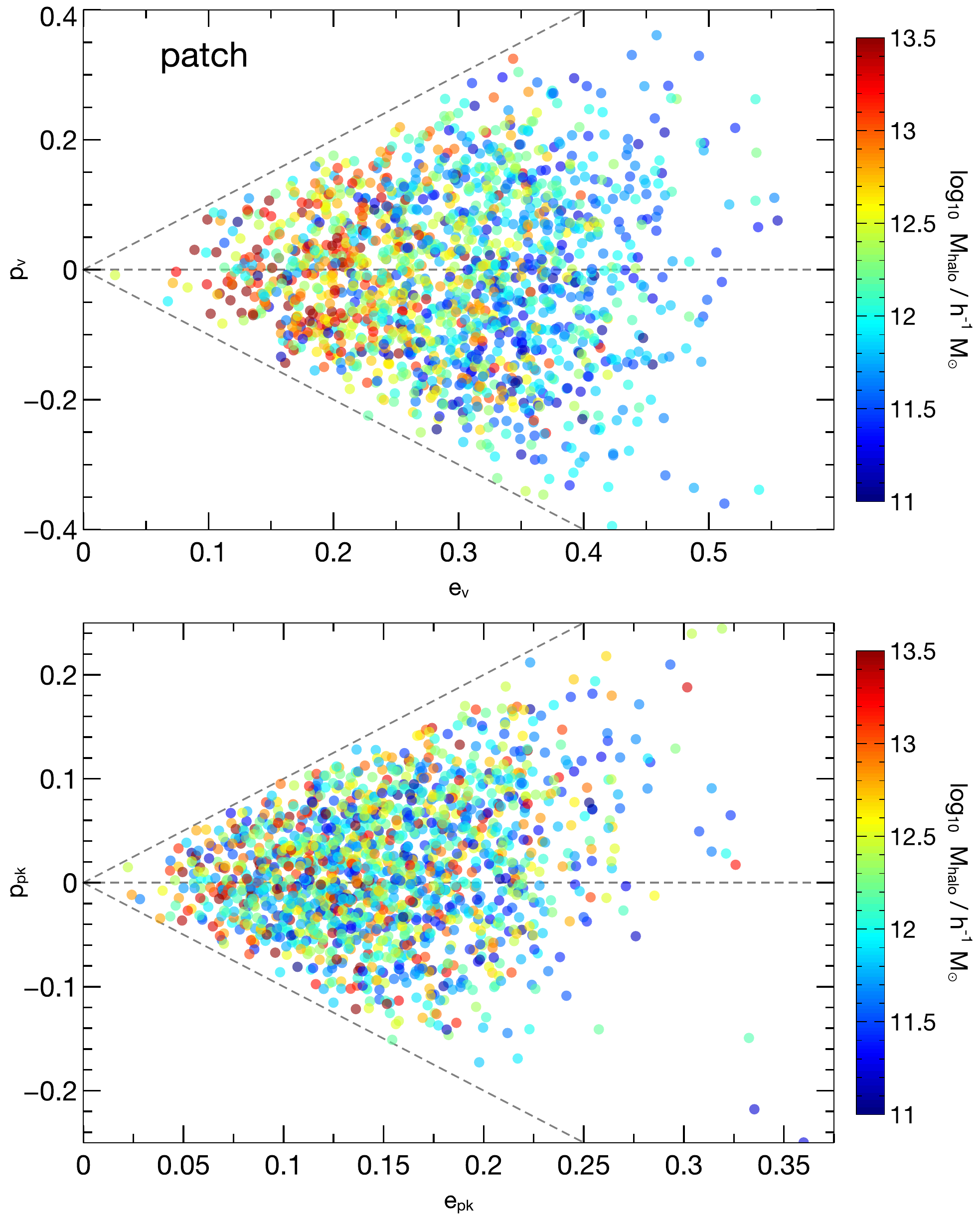}
\end{center}
\caption{(\emph{Top panel}:) The distribution of patch-averaged ellipticity $e_{\rm v}$ (equation~\ref{ev-Y-def}) and prolateness $p_{\rm v}$ (equation~\ref{pv-Z-def}) of the tidal tensor $\p_{ij}\psi$, coloured by halo mass. Both the median value and scatter of $e_{\rm v}$ increases with decreasing halo mass, whereas we find no trend in $p_{\rm v}$ with halo mass. 
(\emph{Bottom panel}:) Corresponding patch-averaged quantities $e_{\rm pk}$ (equation~\ref{epk-y-def}) and $p_{\rm pk}$ (equation~\ref{ppk-z-def}) defined for the density Hessian $\p_{ij}\del$. (Note that the scale on the axes is different from the top panel.) There is a weak preference for low mass objects to have $p_{\rm pk} > 0$, which is consistent with the BBKS results for peak shapes.
}
\label{fig:shear_shape_mass}
\end{figure}

\begin{figure}
\begin{center}
\includegraphics[width=0.4\textwidth]{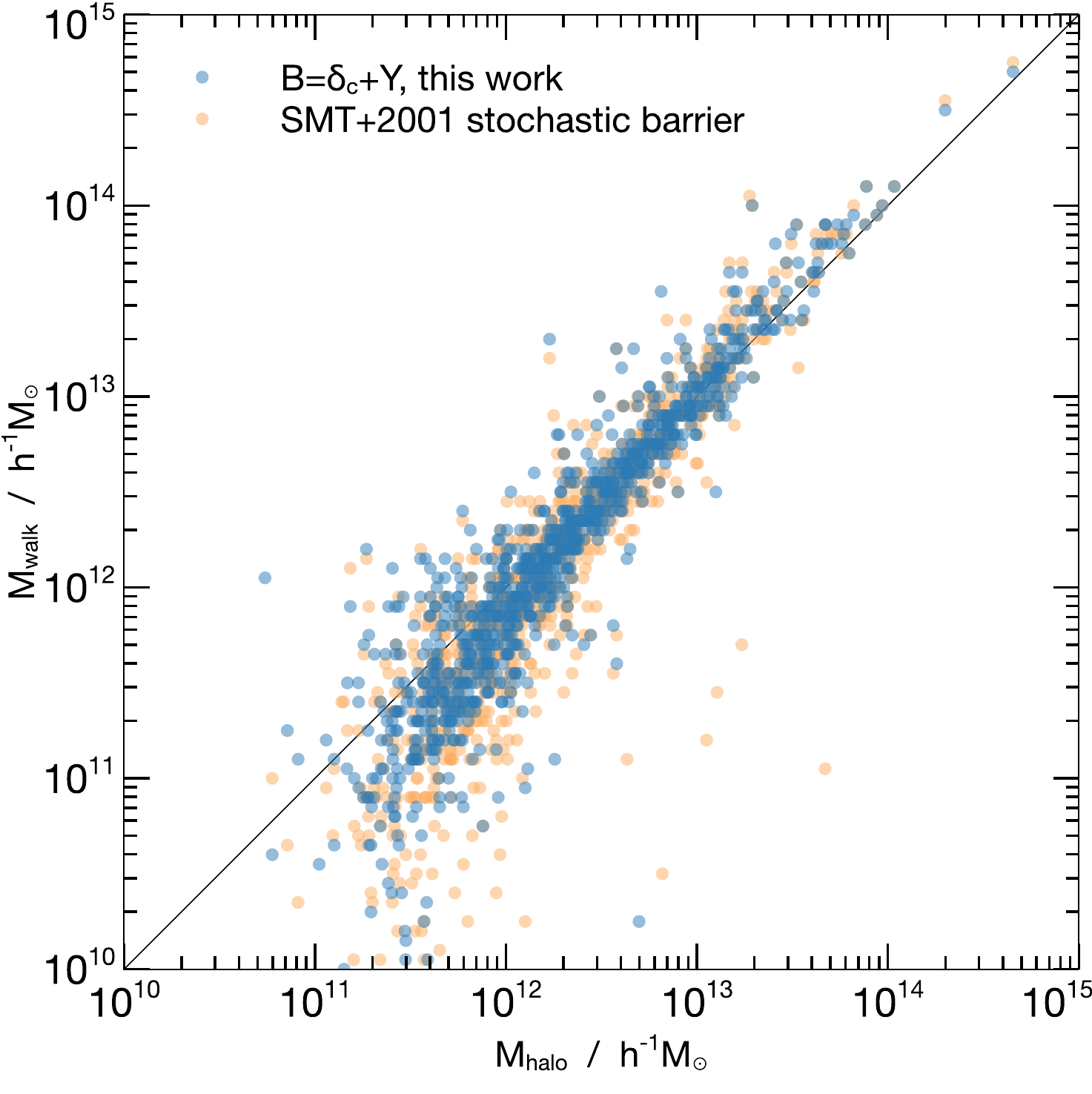}
\end{center}
\caption{Comparing \peak\ masses assigned by the algorithm described in Section~\ref{sec:empirical_walks_method} when using the simple stochastic barrier (equation~\ref{simplebarrier}; blue points) and when using the fully stochastic SMT01 barrier (equation~\ref{SMTbarrier}; orange points). The points show the masses $M_{\rm halo}$ of proto-haloes that could be matched to \peaks, against \peak\ mass $M_{\rm walks}$. In both cases the matched fraction was comparable ($64\%$ when using equation~\ref{simplebarrier} and $62\%$ when using equation~\ref{SMTbarrier}); however, the scatter when using the SMT01 barrier is somewhat larger than when using \eqn{simplebarrier}, especially at low masses, which can be traced to the fact that the SMT01 barrier introduces a dependence of the peak-centred overdensity on prolateness which is not present in the case of the actual proto-haloes.}
\label{fig:smtVssimple}
\end{figure}

Figure~\ref{fig:shear_shape_mass} shows the distributions of ellipticity and prolateness defined using the tidal tensor (top panel) and using the Hessian of the density (bottom panel), coloured by the halo mass in each case. These quantities were defined in equations~\eqref{ev-Y-def}-\eqref{ppk-z-def} above. 

In the top panel we see that low mass haloes have a weak preference for larger values of ellipticity $e_{\rm v}$, while there is no trend of prolateness $p_{\rm v}$ with mass. This is consistent with the results of \citet{lp11} and \citet{dts13} for CDM haloes. In the bottom panel (note the difference in axes scales) we see that there is a large scatter in values of $e_{\rm pk}$ at any mass, while there is a weak trend for $p_{\rm pk} > 0$ at low masses. The latter is consistent with the BBKS results for peak shapes.

We noted earlier that the barrier \eqref{SMTbarrier} associated with ellipsoidal dynamics does not describe the measured densities of the proto-halo patches; in particular, we found no correlation between the measured proto-halo density and prolateness. We have repeated the exercise of finding \peaks\ and matching them with proto-haloes using the full ellipsoidal collapse barrier \eqref{SMTbarrier}. While we find that a similar fraction ($\sim62\%$) of proto-haloes have matching \peaks, the scatter in the assigned masses, especially at low masses, is somewhat larger in this case as compared to using \eqn{simplebarrier}.  Figure~\ref{fig:smtVssimple} compares the mass distributions of matched objects in these two cases. 

\label{lastpage}

\end{document}